\documentclass[cjournal]{IEEEtran}
\usepackage{graphicx}
\usepackage{cite,color}
\usepackage{amsmath,amssymb,amsfonts}
\usepackage{algorithmic}
\usepackage{algorithm,algorithmic}
\usepackage{hyperref}
\hypersetup{
colorlinks= false,
    linkcolor=blue
    }

\def\fbf{{\bf f}}

\def\hbf{{\bf h}}

\def\sbf{{\bf s}}

\def\ubf{{\bf u}}

\def\wbf{{\bf w}}

\def\ybf{{\bf y}}
\def\zbf{{\bf z}}

\def\ybf{{\bf y}}

\def\Hc{{\cal H}}

\def\Mc{{\cal M}}

\def\Oc{{\cal O}}

\def\ie{{\it i.e.,\ \/}}
\def\nn{\nonumber}

\newtheorem{proposition}{Proposition}

\title{Beamforming and Device Selection Design in Federated Learning with Over-the-air Aggregation}

\author{
  \text{Faeze Moradi Kalarde}\textsuperscript{$\star$}\thanks{\textsuperscript{$\star$}\text{University of Toronto}},
  \text{Min Dong}\textsuperscript{$\dagger$}\thanks{\textsuperscript{$\dagger$}\text{Ontario Tech University}},
  \text{Ben Liang}\textsuperscript{$\star$},
  \text{Yahia A. Eldemerdash Ahmed}\textsuperscript{$\ddagger$}\thanks{\textsuperscript{$\ddagger$}\text{Ericsson Canada}},   \text{Ho Ting Cheng}\textsuperscript{$\ddagger$}
  }

\begin{document}

\maketitle

\begin{abstract}
Federated learning (FL) with over-the-air computation can efficiently utilize the communication bandwidth but is susceptible to analog aggregation error. Excluding those devices with weak channel conditions can reduce the aggregation error, but it also limits the amount of local training data for FL, which can reduce the training convergence rate. In this work, we jointly design uplink receiver beamforming and device selection for over-the-air FL over time-varying wireless channels  to maximize the training convergence rate. We reformulate this stochastic optimization problem into a mixed-integer program using an upper bound on the global training loss over communication rounds. We then
propose a Greedy Spatial Device Selection (GSDS) approach, which uses a sequential procedure to select devices based on a measure capturing both the channel strength and the channel correlation to the selected devices. We show that given the selected devices, the receiver beamforming optimization problem is equivalent to downlink single-group multicast beamforming. To reduce the computational complexity, we also propose an Alternating-optimization-based Device Selection and Beamforming (ADSBF) approach,  which solves the receiver beamforming and device selection subproblems alternatingly. In particular, despite the device selection being an integer problem, we are able to develop an efficient algorithm to find its optimal solution.
 Simulation results with real-world image classification demonstrate that our proposed methods achieve faster convergence with significantly lower computational complexity than existing alternatives. Furthermore, although ADSBF shows marginally inferior performance to GSDS, it offers the advantage of lower computational complexity when the number of devices is large.
\end{abstract}

\begin{IEEEkeywords}
Federated learning, over-the-air aggregation, device selection, receiver beamforming.
\end{IEEEkeywords}

\section{Introduction}
\IEEEPARstart{F}{e} derated learning (FL) is an effective distributed machine learning technique that allows multiple devices to collaboratively learn a global model using their local datasets  \cite{zhu2020toward,lim2020federated}. A parameter server needs to aggregate the local model updates from the devices to perform a global model update. However, in wireless FL, information exchange between the devices and the server can create stress on the limited communication resources, especially when a substantial number of devices participate in the process. In such a scenario, the devices may not be able to send their updates simultaneously via conventional orthogonal multiple access over the limited available bandwidth.

In order to improve communication efficiency in wireless FL, analog aggregation of the local models has been proposed \cite{zhu2019broadband}. In this approach, the devices simultaneously transmit their local models using analog modulation over a shared wireless uplink channel, which naturally results in model aggregation at the receiver by superposition. Such over-the-air computation has attracted growing interest due to its advantages of efficient utilization of bandwidth and reduced communication latency over the conventional approach of orthogonal multiple access \cite{amiri2020machine,amiri2020federated,sery2020analog,guo2020analog,zhang2021gradient,zhang2021federated,wang2022online}. Some recent works have also developed real-life prototypes for FL with over-the-air aggregation \cite{PracticalImplementation1, PracticalImplementation2, PracticalImplementation3, PracticalImplementation4}.

However, over-the-air computation is susceptible to noise, which can cause significant aggregation errors that propagate over the FL computation and communication iterations. Furthermore, the quality of aggregation is disproportionately affected by devices with weak channel conditions, since the devices with strong channel conditions have to reduce their transmit power in order to align their transmitted signal amplitude with that of devices experiencing weak channels. This adjustment leads to a lower received signal-to-noise ratio (SNR) \cite{lim2020federated}. Although excluding devices with weak channels can reduce the aggregation error, it can also harm the learning performance as a result of the reduced size of training data. Therefore, we need to carefully design an effective method for device selection that can properly trade off these two effects to improve the overall FL training performance.

Device selection for over-the-air FL was first considered in \cite{zhu2019broadband} and later in \cite{sery2021over} and \cite{ImperfectDownlink}. In \cite{zhu2019broadband}, a distance-based device selection method within a cell was proposed to increase the received SNR at the base station (BS). In \cite{sery2021over}, the authors proposed to only select the devices with strong channel conditions in the learning process to improve the convergence of the model training. However, how to design a proper threshold on the channel strength for device selection was not discussed. In \cite{ImperfectDownlink}, under imperfect downlink channel conditions for model broadcasting, device selection and transmit power at both devices and the server were jointly optimized to minimize the global training loss. All these works assume a single antenna at the server for communication, and thus the multi-antenna receiver processing was not considered.

In current wireless networks, the server is typically equipped with multiple antennas, where beamforming techniques can be applied to enhance the signal strength and reduce the noise in over-the-air computation \cite{zhu2018mimo,jiang2019over}. It was demonstrated in \cite{firstyang2020federated} that the method in \cite{jiang2019over} can be applied to improve FL performance. However, the study in \cite{firstyang2020federated} did not consider device selection. Joint receiver beamforming and device selection was studied in \cite{yang2020federated,Paper2023,liu2021reconfigurable}. In \cite{yang2020federated}, the joint design aimed to maximize the number of selected devices while limiting the communication error by a target threshold, and a difference-of-convex-function (DC) programming method was proposed to solve the joint optimization problem. For the same problem considered in \cite{yang2020federated}, the authors of \cite{Paper2023} introduced a low-complexity method to design the receiver beamforming and device selection jointly. FL via a reflective intelligent surface (RIS)-assisted wireless system was considered in \cite{liu2021reconfigurable}, where device selection, receiver beamforming, and RIS phase shift were jointly considered to minimize an upper bound on the steady-state expected global loss as the training time approaches infinity. In their proposed scheme, the successive convex approximation (SCA) method was used to design receiver beamforming, and Gibbs sampling was used for device selection. 

However, there are some limitations in these existing methods. First, the design and analysis of these works all assume that the channel states remain unchanged for all communication rounds during the entire FL training process, which is unrealistic in practical systems.
Also, for \cite{yang2020federated} and \cite{Paper2023}, the design objective of maximizing the number of selected devices does not directly measure the joint impact of device selection and imperfect communication on FL training performance. The main
challenge of this approach is that it is unclear how to properly set the target threshold for the communication error, which represents the proper trade-off between the communication error and the impact of device selection on the FL training convergence. Although
\cite{liu2021reconfigurable} directly uses the global training loss as the design objective, the proposed Gibbs sampling method has high computational complexity as the number of devices grows, which is undesirable for implementation in practical systems, especially when device selection needs to be updated in each communication round. It is important to design a low-complexity algorithm for device selection and receiver beamforming that effectively improves over-the-air FL training performance.

Given the above issues, in this work, we consider FL with over-the-air aggregation.  Aiming at improving the training convergence rate for FL, we jointly design uplink receiver beamforming and device selection to minimize the global training loss after arbitrary $T$ communication rounds, subject to per-device average transmit power constraint. Unlike the existing works \cite{yang2020federated,Paper2023,liu2021reconfigurable}, we formulate our design problem assuming the channel states between the server and the devices can change over communication rounds, and the device selection and beamforming solutions are computed in each round based on the current channel state information. The main contribution of this paper is summarized as follows:
\begin{itemize}
    \item The formulated joint receiver beamforming and device selection problem is a challenging finite time-horizon stochastic optimization problem. By analyzing the training procedure, we obtain an upper bound for the global training loss function after $T$ communication rounds. To improve the convergence rate of FL, we design receiver beamforming and device selection to minimize this upper bound on the global loss.
    \item The reformulated joint optimization problem is a mixed-integer programming problem that presents significant challenges. 
    We first propose a Greedy Spatial Device Selection (GSDS) approach to obtain a solution. GSDS uses a greedy method to select the devices and then solves the corresponding beamforming optimization problem among the selected devices. In particular, GSDS uses a sequential procedure to add devices to the set of selected devices based on a metric that measures the channel strength and the channel correlation to the selected devices. Given the selected devices, we show that the optimization problem with respect to (w.r.t.) receiver beamforming is equivalent to downlink single-group multicast beamforming, for which we apply an SCA method to obtain a solution. The overall computational complexity of GSDS is shown to grow with the number of devices $M$ as $\mathcal{O}(M^3)$.
    \item To reduce the computational complexity, we further devise an Alternating-optimization-based Device Selection and Beamforming (ADSBF) approach. ADSBF employs the alternating optimization technique to break the joint optimization problem into two subproblems w.r.t. receiver beamforming and device selection and solve them alternatingly. We show that the receiver beamforming subproblem is the same as that in GSDS and can be solved in the same manner. For the device selection subproblem, despite being an integer problem that generally is difficult to solve, we are able to develop an efficient algorithm in our problem to find the optimal solution with computational complexity $\mathcal{O}(M\text{log}(M))$.  

    \item We test the proposed GSDS and ADSBF with real-world image classification tasks using MNIST and CIFAR-10 datasets. Our simulation results demonstrate that both GSDS and ADSBF outperform existing methods for beamforming and device selection design in terms of the training convergence rate. They also have significantly lower computational complexity compared with the existing methods. This shows the effectiveness of our proposed approaches in providing a proper trade-off between the impact of noisy communication and the amount of training data.
    Between the two approaches, GSDS performs slightly better than ADSBF as it leads to faster convergence. However, the run time of ADSBF is significantly lower than that of GSDS when the number of devices is large.
\end{itemize}

The rest of this paper is organized as follows. In Section II, we present the system model and problem formulation for over-the-air FL. In Section III, we reformulate the problem via training convergence analysis. In Section IV, we propose GSDS and ADSBF approaches to obtain the solution. The simulation result is provided in Section V, followed by the conclusion in Section VI.

\allowdisplaybreaks

\section{System Model and Problem Formulation}

\subsection{FL System }

We consider a wireless network consisting of a server and $M$ local devices. The set of the devices is denoted by $\mathcal{M}$. Device $m$ has a local training dataset of size $K_m$ and denoted by $\mathcal{D}_m=\{(\mathbf{x}_{m,k}, y_{m,k}): 1\le k \le K_m \}$, where $\mathbf{x}_{m,k} \in \mathbb{R}^b$ is the $k$-th data feature vector, and $ y_{m,k}$ is its corresponding label. The devices aim to collaboratively train a global model at the server that can predict the true labels of data feature vectors from all devices while keeping their local datasets private. The empirical local training loss function at device $m$ is defined as
        \begin{align}\label{eq1}
            F_m(\mathbf{w}; \mathcal{D}_m)  \triangleq \frac{1}{K_m} \sum \limits_{k=1}^{K_m} l(\mathbf{w}; \mathbf{x}_{m,k} , y_{m,k}), 
        \end{align}
where $\mathbf{w} \in \mathbb{R}^D$ is the global model parameter vector, and $l(\cdot)$ is the sample-wise training loss function associated with each data sample. The global training loss function is defined as the weighted sum of the local loss functions over all devices, given by
    \begin{align}\label{eq2}
    F(\mathbf{w})= \frac{1}{K}\sum \limits_{m=1}^M K_m F_m(\mathbf{w}; \mathcal{D}_m ), 
    \end{align}
where $K= \sum_{m=1}^M K_m $ is the total number of training samples over all devices. 

We follow the general Federated Stochastic Gradient Descent (FedSGD) approach for the iterative model training in FL, where the server updates the global model parameters based on an aggregation of the gradients of all devices’ local loss functions \cite{mcmahan2017communication}. The learning objective is to find the optimal global model $\mathbf{w}^\star$ that minimizes the global training loss function $F(\mathbf{w})$. We call each iteration of the global model update a communication round. In communication round $t$, the following steps are performed:
\begin{enumerate}
    \item \textbf{Device selection:} The server selects a subset of devices to contribute to the training of the model. The set of selected devices in round $t$ is denoted by $\mathcal{M}_t^\text{s} \subseteq  \mathcal{M}$.
   \item \textbf{Downlink phase:} The server broadcasts the model parameter vector $\mathbf{w}_t$ to all devices and notifies the selected devices.
    \item  \textbf{Local gradient computation:} Each selected device $m$ computes the gradient of its local loss function, given by
        $\mathbf{g}_{m,t} = \nabla F_m(\mathbf{w}_t; \mathcal{D}_m)$,
    where $\nabla F_m(\mathbf{w}_t; \mathcal{D}_m)$ is the gradient of $F_m(\cdot)$ at $\mathbf{w}_t$.
    \item \textbf{Uplink phase:} The selected devices send their local gradients to the server through their uplink wireless channels.
    \item  \textbf{Global model update:} The server computes a weighted aggregation of local gradients to update the global model. In the ideal scenario where the local gradients can be received at the server accurately, the weighted aggregation $\mathbf{r}_t \triangleq \sum_{m \in \mathcal{M}_t^\text{s}} K_m \mathbf{g}_{m,t} \in \mathbb{R}^D$ is used to update $\wbf_t$. However, in practice, the received complex base-band signal processed at the server $\mathbf{\hat r}_t \in \mathbb{C}^D$ is imperfect due to the noisy communication channels. Thus, the server updates the global model as
    \begin{align}\label{eq5}
                \mathbf{w}_{t+1} = \mathbf{w}_t - \frac{\lambda}{ \sum_{m \in \mathcal{M}_t^\text{s}} K_m } \Re[{\mathbf{\hat r}_t}],
     \end{align}
    where $\lambda$ is the learning rate, and $\Re[\cdot]$ represents the real part of a complex variable.
\end{enumerate}

\subsection{FL with Over-the-Air Analog Aggregation}
\begin{figure*}[t]
\centerline{\includegraphics[width=6in]{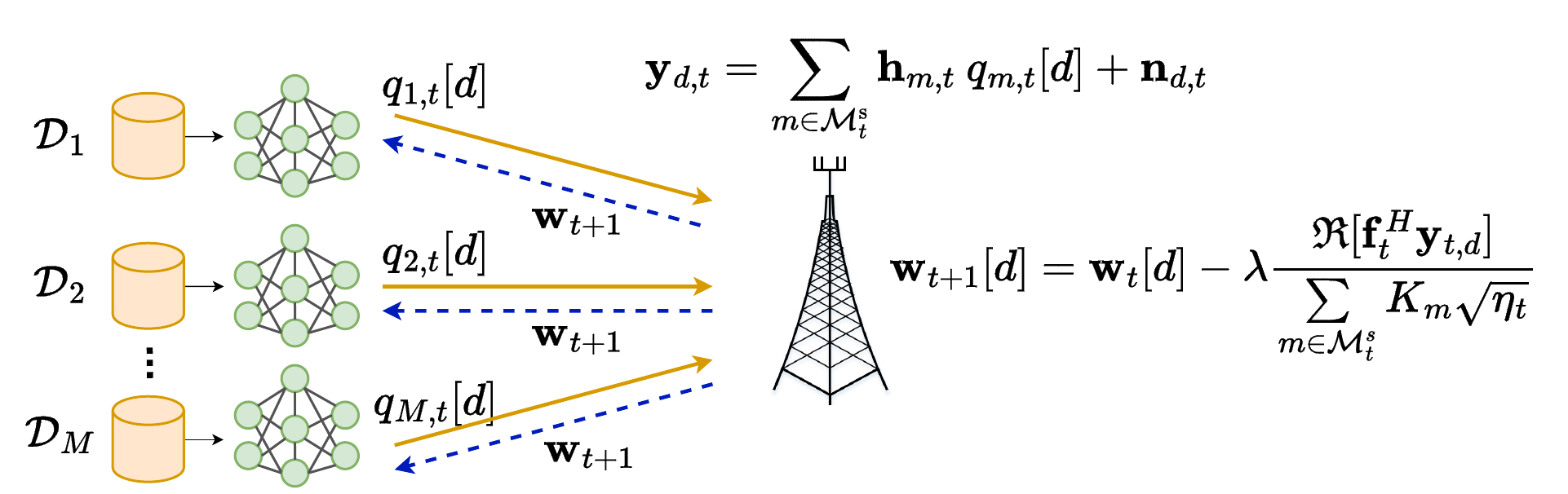}}
\caption{An illustration of federated learning with over-the-air aggregation.\label{SystemFig}}
\end{figure*}

We assume the server is equipped with $N$ antennas, and each device has a single antenna. The uplink channel between device $m$ and the server in communication round $t$ is denoted by $\mathbf{h}_{m,t} \in \mathbb{C}^{N}$. We consider over-the-air analog aggregation over the multiple access channel to efficiently obtain the aggregated local gradients at the server \cite{zhu2019broadband}. The FL system with over-the-air aggregation is shown in Fig. \ref{SystemFig}. In each communication round, the devices send their local gradients to the server simultaneously over the same frequency band. Due to the superposition of the transmitted signals, the server receives the weighted sum of local gradients. 

At each communication round $t$, the local gradient of each selected device is transmitted over $D$ channel uses (or symbol durations). Specifically, the local gradient vector $\mathbf{g}_{m,t}$ at the selected device 
$m$ is first normalized by its local gradient normalization scalar $v_{m,t}$, and then adjusted by the transmit weight $a_{m,t} \in C$ for transmission. Thus, the transmitted signal for the $d$-th entry of the local gradient $g_{m,t}[d]$, which is denoted by $q_{m,t}[d]$, is given by
\begin{align}\label{TrnamsitedSignal}
    q_{m,t}[d] = a_{m,t} \frac{g_{m,t}[d]}{v_{m,t}}.
\end{align}
To facilitate the receiver processing at the server, each selected device $m$ sends its local gradient normalization scalar $v_{m,t}=\frac{\|\mathbf{g}_{m,t}\|}{\sqrt{D}}$ via the uplink signaling channel to facilitate the receiver processing. We assume the signaling channel is a separate digital channel and the reception is perfect. 

Let $\mathbf{q}_{m,t}= [q_{m,t}[1], ..., q_{m,t}[D]]^T$ denote the corresponding transmitted signal vector for $\mathbf{g}_{m,t}$. From \eqref{TrnamsitedSignal}, the average transmit power used to send each entry in $\mathbf{g}_{m,t}$ from device $m$ in communication round $t$ is $\frac{\|\mathbf{q}_{m,t}\|^2}{D} = |a_{m,t}|^2$.
The average transmit power is subject to the maximum average transmit power limit $P_0$: $|a_{m,t}|^2 \le P_0$, $\forall m, \forall t$.

The corresponding received signal at the server for the $d$-th channel use is given by
        \begin{align}\label{eq10}
            \mathbf{y}_{d,t}= \sum_{m \in \mathcal{M}_t^\text{s} } \mathbf{h}_{m,t} q_{m,t}[d]+\mathbf{n}_{d,t},
        \end{align}
where $\mathbf{n}_{d,t} \sim \mathcal{CN}(\mathbf{0}, \sigma_n^2 \mathbf{I})$ is the receiver additive white Gaussian noise for the $d$-th channel use and is i.i.d. over $t$. 

The server applies receive beamforming to process the received signal. Let $\mathbf{f}_t \in \mathbb{C}^N$ denote the receive beamforming vector in communication round $t$ with $\|\mathbf{f}_t\| =1$, and let $\eta_t \in \mathbb{R}^+$ denote the receive scaling factor. The post-processed received signal for $\mathbf{y}_{d,t}$ is given by
\begin{align}\label{eq11}
  \hat{r}_t[d]  & =  \frac{\mathbf{f}_t^H \mathbf{y}_{d,t}}{\sqrt{\eta_t}} \nonumber \\ & =  \frac{1}{\sqrt{\eta_t}}\Big(\sum_{m \in \mathcal{M}_t^\text{s}} \mathbf{f}_t^H \mathbf{h}_{m,t} a_{m,t}\frac{g_{m,t}[d]}{v_{m,t}}+ \mathbf{f}_t^H \mathbf{n}_{d,t} \Big).
\end{align}
Let $\mathbf{\hat{r}}_t \triangleq [\hat{r}_t[1],...,\hat{r}_t[D]]^T$. The server uses the real part of the post-processed signal $\Re[{\mathbf{\hat r}_t}]$, which contains the received local gradients from the selected devices to update the global model $\wbf_t$ based on \eqref{eq5}.\footnote{Note that for more efficient transmission, $\mathbf{g}_{m,t}$ can be sent via complex signals using both the real and imaginary parts of the signal. The receiver can perform post-processing to extract information from the real and imaginary parts of the received signal. In this work, we only consider information being sent using a real-valued signal for simplicity. This will not change the fundamental process developed subsequently.}
   
\begin{figure*}[t]
\centerline{\includegraphics[width=5.5in]{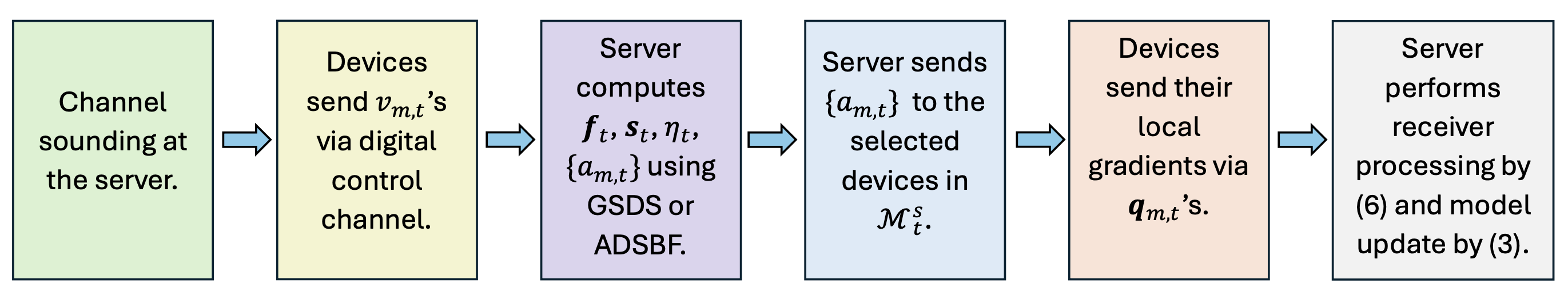}}
\caption{An illustration of the sequential operations performed within communication round t.}\label{Procedure}
\end{figure*}

We describe the sequential operations performed within each communication round in Fig.~\ref{Procedure} and summarize the key notations used throughout this work in Table~\ref{tab1}. 

\begin{table}[!t]
\caption{Notations}
\label{table1}
\setlength{\tabcolsep}{3pt}
\begin{tabular}{|p{25pt}||p{200pt}|}
\hline
\textbf{Symbol}& 
\textbf{Explanation}\\
\hline
$\mathbf{w}_t $ & Model parameter vector in round $t$ \\
\hline
$F_m(\cdot) $ & Local loss function of device $m$ \\
\hline
$ F(\cdot)$ & Global loss function \\
\hline
$\mathcal{D}_m$ & Local dataset of device $m$ \\
\hline
$\mathbf{g}_{m,t}$ & Gradient of local loss function of device $m$ in round $t$   \\
\hline
$K_m$ & Size of local dataset of device $m$ \\
\hline
$K$ & Total number of data samples over all devices \\
\hline
$\mathbf{h}_{m,t}$ & Channel condition of device $m$ in round $t$\\
\hline
$\mathbf{f}_t$ & Receive beamforming in round $t$ \\
\hline
$\eta_t$ & Receive scaling factor in round $t$\\
\hline
$ \mathbf{s}_t$ & Device selection vector in round $t$\\
\hline
$ \mathcal{M}$ & Set of all devices \\
\hline
$ \mathcal{M}_t^\text{s}$ & Device selection set in round $t$\\
\hline
$a_{m,t}$ & Transmit weight of device $m$ in round $t$ \\
\hline
$M$ & Number of devices \\
\hline
$N$ & Number of antennas at server\\
\hline
$P_0$ & Average transmit power limit \\
\hline
$\mathbf{n}_{d,t}$ & Additive noise vector in $d$-th channel use in $t$-th round\\
\hline
$\sigma_n^2$ & The variance of each entry of the noise vector\\
\hline
$\lambda$ & Learning rate \\
\hline
$I_{\text{max}}$ & Maximum number of SCA iterations\\
\hline
$J_{\text{max}}$ & Maximum number of ADSBF iterations\\
\hline
\end{tabular}
\label{tab1}
\end{table}

\subsection{Problem Formulation}

Since learning efficiency is crucial for FL, we aim to maximize the training convergence rate. As such, our objective is to minimize the expected global loss function after $T$ communication rounds by jointly optimizing the device selection $ \mathcal{M}_t^\text{s}$, the devices transmit weights $\{a_{m,t}\}$, and the receiver processing (beamforming vector $
\mathbf{f}_t$ and scaling factor $\eta_t$). This optimization problem is formulated as follows:
\begin{align}\label{eq33}
    \min_{\{\mathbf{f}_t, \mathbf{s}_t , \eta_t, \{a_{m,t}\} \}_{t=0}^{T-1}} \quad & \mathbb{E}[F(\mathbf{w}_T)]\\
    \textrm{s.t.} \quad &
    |a_{m,t}|^2 \le P_0, \: \forall m , \forall t , \label{TranmsitPowerCons}\\
    & \|\mathbf{f}_t\| =1, \forall t, \\ & 
    \eta_t > 0, \forall t, \\ &
    \mathbf{s}_t \in \{0,1\}^M, \forall t, 
    \end{align}
where $\mathbb{E}[\cdot]$ is the expectation taken over the receiver noise, and $\mathbf{s}_t$ is the binary device selection vector at round $t$, with its $m$-th entry $s_{t}[m]=1$ indicating device $m$ is selected to participate the model updating in  communication round $t$ and $0$ otherwise. Note that $\mathbf{s}_t$ and $\mathcal{M}_t^\text{s}$ convey the same information, and we have $\mathcal{M}_t^\text{s} = \{ m: {s}_t[m]=1, m \in \mathcal{M}\}$. 

\section{Problem Reformulation Based on Global Training Loss Analysis}

Problem \eqref{eq33} is a finite time horizon stochastic optimization problem, and the global loss function in the objective is not an explicit function of the optimization variables. Furthermore, it requires the knowledge of channel states, $\{\hbf_{m,t}\}$ for
$t = 0, ..., T-1$, and therefore, it cannot be solved in an online fashion as the channel state in the future is unavailable. In addition,  problem \eqref{eq33} is a mixed integer program as it contains a binary variable $\mathbf{s}_t$. To tackle this challenging problem, we consider a more tractable upper bound on the loss function through training loss convergence analysis and propose algorithms to minimize this upper bound. 

\subsection{Upper Bound on the Global Training Loss}
To analyze the expression of $F(\mathbf{w}_T)$, we rewrite the global model update at the server in \eqref{eq5} as 
     \begin{align}\label{eq12}
         \mathbf{w}_{t+1} = \mathbf{w}_t - \lambda (\nabla F(\mathbf{w}_t)+ \mathbf{e}_t),
     \end{align}
where $\nabla F(\mathbf{w}_t)$ is the gradient of global loss function at $\mathbf{w}_t$, and $\mathbf{e}_t \in \mathbb{R}^D$ is the error vector representing the deviation of the updating direction from $\nabla F(\mathbf{w}_t)$. Based on \eqref{eq5}, the error vector $\mathbf{e}_t$ can be expressed as follows:
\begin{align}\label{ErrorSplitting}
         \mathbf{e}_t &= \frac{\Re{[\hat{\bold{r}}_{t}]}}{\sum_{m \in \mathcal{M}_t^\text{s}}K_{m}}-\nabla F(\mathbf{w}_{t}) \nonumber \\ 
&=    \underbrace{\frac{\Re{[\hat{\bold{r}}_{t}]}}{\sum\limits_{m \in \mathcal{M}_t^\text{s}}K_{m}}- \frac{\bold{r}_{t}}{\sum\limits_{m \in \mathcal{M}_t^\text{s}}K_{m}}}_\mathrm{\triangleq \mathbf{e_{1, t}}}+    \underbrace{ \frac{\bold{r}_{t}}{\sum\limits_{m \in \mathcal{M}_t^\text{s}} K_{m}}- \nabla F(\mathbf{w}_{t})}_\mathrm{\triangleq \mathbf{e_{2, t}}}, 
\end{align}
where the first component of the error $\mathbf{e}_{1,t}$ arises from the difference between the accurate aggregation $\mathbf{r}_t$ and its server-estimated counterpart $\Re{[\hat{\bold{r}}_{t}]}$, due to imperfect communication, stemming from receiver noise and analog aggregation across wireless channels. The second component of the error $\mathbf{e}_{2,t}$ is due to the subset selection of devices. When all devices are chosen, this term disappears. 
Note that  the gradient error $\mathbf{e}_{t}$ in \eqref{ErrorSplitting} depends on the device selection set $\mathcal{M}_t^\text{s}$, transmit weights of devices $\{a_{m,t}\}$, and receiver processing ($\mathbf{f}_t$, $\eta_t$) through $\hat{\bold{r}}_{t}$, as shown in \eqref{eq11}.

From \eqref{eq11}, $\hat{\bold{r}}_{t}$ is a function of the selected devices, the transmit weight $\{a_{m,t}\}$ at these devices, the receive beamforming $\mathbf{f}_t$, and the receive scaling $\eta_t$ at the server. Given selected devices set $\Mc_t^{s}$ and receive beamforming $\mathbf{f}_t$, we want to optimize the receive scaling $\eta_t$ and transmit weight $\{a_{m,t}\}$ to minimize the expected error $\mathbb{E}[\|\mathbf{e}_{1,t}\|^2]$ under the transmit power constraint \eqref{TranmsitPowerCons}, which is given by
\begin{align}\label{ErrorMinimization}
    \min_{\eta_t, \{a_{m,t}\}} \quad & \mathbb{E}[\|\mathbf{e}_{1,t}\|^2]\\
    \textrm{s.t.} \quad &
    |a_{m,t}|^2 \le P_0, \: \forall m,
    \\ & \eta_t > 0.
    \end{align}
However, the optimal solution to the error minimization problem \eqref{ErrorMinimization} is not straightforward to obtain. Thus, we adopt a sub-optimal solution to this problem, which has been considered by the existing works \cite{yang2020federated,liu2021reconfigurable, Paper2023 }:
\begin{align}\label{sub-optimal-sol-1}
         & \eta_t  = \min\limits_{m\in \mathcal{M}_t^\text{s}}\frac{ P_{0} \lvert\mathbf{f}_t^{H} \mathbf{h}_{m,t}\rvert ^{2}}{K_{m}^{2}v_{m,t}^{2}},
\end{align}
\begin{align}\label{sub-optimal-sol-2}
         a_{m,t}= \frac{\sqrt{\eta_t} K_m  v_{m,t} }{\mathbf{f}_t^H \mathbf{h}_{m,t} }.
    \end{align}
Substituting the above expressions into \eqref{ErrorSplitting}, we can analyze the impact of $\mathbf{e}_t$ on the expected global loss function based on \eqref{eq12}.  An error expression similar to \eqref{ErrorSplitting} is analyzed in \cite{liu2021reconfigurable} for an RIS-assisted FL system, where an upper bound on the expected global loss function $\mathbb{E}[F(\mathbf{w}_{t+1})]$ is derived. We can apply this upper bound to our problem straightforwardly by replacing the RIS channel with our channel model. The upper bound is derived based on the following assumptions on the global
loss function $F(\mathbf{w})$, which are common in the stochastic optimization literature \cite{friedlander2012hybrid}:

\begin{enumerate}
\item[{\bf A1.}] The global loss function $F(\mathbf{w})$ is twice continuously differentiable.
\item[{\bf A2.}]  The global loss function $F(\mathbf{w})$ is $\mu$-strongly convex:  $\forall \mathbf{w}, \mathbf{w}^\prime \in \mathbb{R}^D$,
    \begin{align}\label{StrongConvexity}
\hspace*{-1em}        F(\mathbf{w})\geq  F(\mathbf{w}^\prime) + (\mathbf{w}-\mathbf{w}^\prime)^{T} \nabla F(\mathbf{w}^\prime)+\frac{\mu}{2}\|\mathbf{w}-\mathbf{w}^\prime\|^{2}_{2}. 
    \end{align}
\item[{\bf A3.}]  
 The gradient of the global loss function is Lipschitz continuous with a positive Lipschitz constant $L$: $\forall \mathbf{w}, \mathbf{w}^{\prime} \in \mathbb{R}^D$,
    \begin{align}\label{LipschitzContinuity}
        \|\nabla F(\mathbf{w})-\nabla F({\mathbf{w}}^\prime)\|_{2} \le   L \|\mathbf{w}-\mathbf{w}^\prime\|_{2}.
     \end{align}
\item[{\bf A4.}] The gradient of sample-wise training loss function is upper bounded: $\exists \; \alpha_1 \ge 0$ and $\alpha_2 \ge 1$, s.t.
    \begin{align}\label{eq18}
\hspace*{-1em}        \|\nabla l(\mathbf{w}_{t} ; \mathbf{x}_{k},y_{k})\|^{2}_{2} \le \alpha_{1} \!+ \! \alpha_{2} \|\nabla F (\mathbf{w}_{t})\|^{2}_{2}, \forall k , \! \forall t.
 \end{align}
\end{enumerate}
By substituting the expressions of $\eta_t$ and $\{a_{m,t}\}$ in \eqref{sub-optimal-sol-1} and \eqref{sub-optimal-sol-2} into \eqref{eq11} and setting the learning rate $\lambda = \frac{1}{L}$ in \eqref{eq5},  and based on Assumptions A1-A4, the expected difference between the global loss function at round $(t+1)$ and the optimal loss is bounded by\cite{liu2021reconfigurable}
\begin{align}\label{eq19}
\mathbb{E} [ & F(\mathbf{w}_{t+1})-F(\mathbf{w}^\star) ]  \nonumber \\ &
\leq \psi_t \mathbb{E}[F(\mathbf{w}_t)- F(\mathbf{w}^\star)]+
        \frac{\alpha_1}{L} d(\mathbf{f}_t, \mathbf{s}_t;\mathcal{H}_t),
\end{align}
where $\psi_t \triangleq 1- \frac{\mu}{L} \big( 1-2 \alpha_2 d(\mathbf{f}_t, \mathbf{s}_t;\mathcal{H}_t) \big)$, $\mathcal{H}_t \triangleq \{\hbf_{m,t}: m \in \mathcal{M} \}$,
\begin{align}\label{d_func}
d(\mathbf{f}_t, \mathbf{s}_t;\mathcal{H}_t)\triangleq& \frac{4}{K^2}\Big(\sum_{m=1}^M(1-s_{t}[m])K_m\Big)^2  \nonumber  \\ & + \frac{\sigma_n^2}{P_{0}(\sum_{m=1}^M s_{t}[m] K_m)^2}\max_{1\le m \le M}\frac{s_{t}[m] K_m^2}{\lvert\mathbf{f}_t^{H}\mathbf{h}_{m,t}\rvert^2},
\end{align}
and $\alpha_1$ and $\alpha_2$ are given in \eqref{eq18}.

Let $\mathbf{w}_0$ be the initial model parameter vector. Applying the bound in \eqref{eq19} to $\mathbb{E} [F(\mathbf{w}_{t+1})-F(\mathbf{w}^\star) ] $ for $t = 0,\ldots, T-1$, we have the following upper bound after $T$ communication  rounds:
  \begin{align} \label{eq34}
 &      \mathbb{E} [F(\mathbf{w}_{T})-F(\mathbf{w}^\star) ] \leq \!\Big(\prod_{t=0}^{T-1}\psi_t\Big)  \mathbb{E}[F(\mathbf{w}_0)- F(\mathbf{w}^\star)] \nonumber\\
& +        \frac{\alpha_1}{L} \Big( \sum \limits_{t=0}^{T-2}
        \Big(\!\!\prod_{\tau =t+1}^{T-1}\!\! \psi_\tau\Big) d(\mathbf{f}_{t}, \mathbf{s}_{t};\mathcal{H}_t)+d(\mathbf{f}_{T-1}, \mathbf{s}_{T-1};\mathcal{H}_{T-1})\! \Big).
    \end{align} 

\subsection{Problem Reformulation via Training Loss Upper Bound}
Note that minimizing $\mathbb{E}[F(\mathbf{w}_T)]$ in \eqref{eq33} is equivalent to minimizing $\mathbb{E} [ F(\mathbf{w}_{T})-F(\mathbf{w}^\star) ]$. However, $\mathbb{E} [ F(\mathbf{w}_{T})-F(\mathbf{w}^\star) ]$ is difficult to optimize directly. Instead, we minimize its upper bound in \eqref{eq34}. Note that since $\psi_t$ is an increasing function of $d(\mathbf{f}_t, \mathbf{s}_t;\mathcal{H}_t)$, the upper bound in \eqref{eq34} is an increasing function of $d(\mathbf{f}_t, \mathbf{s}_t;\mathcal{H}_t)$, $t=0,\ldots, {T-1}$. Thus, to minimize the upper bound, it is sufficient to minimize $d(\mathbf{f}_t, \mathbf{s}_t;\mathcal{H}_t)$ at each round $t$ w.r.t. $(\mathbf{f}_t,\mathbf{s}_t)$. This joint device selection and receiver beamforming optimization problem is given below, where we drop subscript $t$ from the problem for notation simplicity:
\begin{align}\label{eq22}
    \min_{\mathbf{f}, \mathbf{s}} \quad & d(\mathbf{f}, \mathbf{s;\mathcal{H}}) \\ 
    \textrm{s.t.} &\quad \|\mathbf{f}\| =1, \\& \quad \mathbf{s} \in \{0,1\}^M .
    \end{align}

\section{Joint Device Selection and Receiver Beamforming with Analog Aggregation} \label{Design}
  
The joint device selection and receive beamforming problem \eqref{eq22} is a mixed-integer program and is challenging to solve. Below, we propose two approaches to find a solution. The first approach uses a greedy method to select the devices based on their channel strength and correlation and then solves the corresponding beamforming optimization problem. To reduce the computational complexity, we further propose an alternating-optimization-based approach, where we devise an efficient low-complexity algorithm for the sub-problem of device selection.

\subsection{Greedy Spatial Device Selection (GSDS) Approach} \label{GSDS}

From \eqref{d_func}, we note that the channels of the selected devices, in both their strength and correlation among each other, can affect the value of $d(\mathbf{f}, \mathbf{s;\mathcal{H}})$. Specifically, $d(\mathbf{f}, \mathbf{s;\mathcal{H}})$ is a decreasing function of the channel strength of each selected device. Furthermore, since the same receive beamforming vector $\mathbf{f}$ applies to the channels of all selected devices, having highly correlated channels among these devices improves the minimum beamforming gain among them, which leads to a reduced value of $d(\mathbf{f}, \mathbf{s;\mathcal{H}})$. Based on these two factors, we propose a Greedy Spatial Device Selection (GSDS) approach. It uses a sequential procedure to add a device to the set of selected devices. We use a metric to measure channel strength and its correlation to the set of selected devices. In each step, the device with the maximum metric value is added to the set of selected devices, and then beamforming optimization is performed for \eqref{eq22} under this new set of selected devices. Since the beamforming optimization is performed in each step, we first describe this beamforming design problem, and then detail the device selection procedure for GSDS.

\subsubsection{Receiver Beamforming Design Given Device Selection} \label{GSDS-sub1}
Assume the current set of selected devices is given by $\mathbf{s}$ (or equivalently $\Mc^\text{s}$). Problem \eqref{eq22} is now reduced to the receiver beamforming optimization problem w.r.t. $\mathbf{f}$ at the server. Since the first term of $d(\mathbf{f}, \mathbf{s;\mathcal{H}})$ in \eqref{d_func}  is  a function of  $\mathbf{s}$ only, with given $\mathbf{s}$,   problem \eqref{eq22} is equivalent to
    \begin{align}\label{eq25}
    \min_{\mathbf{f}}  & \max_{m \in \Mc^\text{s}}\frac{ K_m^2}{|\mathbf{f}^{H}\mathbf{h}_m|^2} \\ &    \textrm{s.t.} \quad   \|\mathbf{f}\| =1.
    \end{align}
By introducing the auxiliary variable $c$, we can further rewrite the min-max problem (\ref{eq25}) as its equivalent epigraph form as
    \begin{align}\label{jadid}
    \underset{\mathbf{f}, c}{\text{min}} \quad & c  \\
    \textrm{s.t.} \quad & \frac{ K_m^2}{|\mathbf{f}^{H}\mathbf{h}_m|^2}\le c, \forall  m \in \Mc^\text{s}, \\
&    \|\mathbf{f}\| =1. \label{uninorm_f}
    \end{align}
 Let  $\tilde{\fbf}\triangleq \sqrt{c} \mathbf{f}$. We can directly optimize $\tilde{\fbf}$ instead of $\fbf$ and $c$ separately in   problem (\ref{jadid}). In this case, constraint \eqref{uninorm_f} can be dropped, and the objective function is replaced by $\|\tilde{\fbf}\|^2$. Thus, we further simplify problem (\ref{jadid})  and transform it into the following final equivalent problem: 
\begin{align}\label{eq27}
    \min_{\tilde{\fbf} }\ \ & \|\tilde{\fbf} \|^2 & \\
    \textrm{s.t.} \quad & |\tilde{\fbf}^H\mathbf{h}_m|^2 \ge K_m^2, \forall   m \in \Mc^\text{s}.
\end{align}

After the above transformations, we arrive at our final problem  (\ref{eq27}) that is in fact equivalent to a single-group downlink multicast beamforming quality-of-service (QoS) problem \cite{Sidiropoulos&etal:TSP2006,dong2020multi}:  The BS  transmits a common message to all devices in $\Mc^\text{s}$  using the multicast beamforming vector $\tilde \fbf$, which is optimized to minimize the transmit power while meeting each device's SNR target. In our problem   (\ref{eq27}), the SINR target is $K_m^2$  for each device $m$. The multicast beamforming design problem has been well studied in the literature \cite{Sidiropoulos&etal:TSP2006,Tranetal:SPL14, dong2020multi,chongTSP23}. It is generally an NP-had problem. Nonetheless, effective and efficient algorithms have been proposed in the literature to find a close-to-optimal solution \cite{dong2020multi,chongTSP23}. We adopt the SCA method to solve problem (\ref{eq27}), which is guaranteed to converge to a stationary point  \cite{dong2020multi}. Once  $\tilde{\fbf}$ is obtained, the receive beamforming vector $\fbf$ can be readily computed as $\mathbf{f} = {\tilde{\mathbf{f}}}/{\| \tilde{\mathbf{f}}\|}$.

\subsubsection{Greedy Selection of Devices} \label{GSDS-sub2}
As shown in the above problem \eqref{eq27}, we note that the receiver beamforming optimization essentially is a single-group multicast beamforming problem.
Since the receiver beamforming vector is applied to all device channels, the worst received SNR among devices improves if all devices have good channel conditions and similar channel directions. Based on these heuristics, we propose our greedy device selection process in GSDS. 

\begin{algorithm}[t]
\begin{algorithmic}[1]
\STATE \textbf{Input:}  $\{\mathbf{h}_m, K_m: m \in \Mc\}$.
\STATE \textbf{Initialization:} $\Mc^s_1=\{m_1\}$, where $m_1=\mathop{\arg\max}_{1\le m\le M}\| \hbf_m\|$.

\FOR{$i = 2, ..., M $}

\STATE Compute $\text{Proj}_{\Hc^s_{i-1}}\!\!(\hbf_m)$, for $m \in\Mc\backslash \Mc^s_{i-1}$. 
\STATE Select device 
\begin{align*}
 m_i = \mathop{\arg\max}_{m\in \Mc\backslash \Mc^s_{i-1}}\|\text{Proj}_{\Hc^s_{i-1}}\!\!(\hbf_m)\|.   
\end{align*}
\STATE Set $\Mc^s_i =  \Mc^s_{i-1} \cup  \{m_i\}$.  
  
\STATE Compute receiver beamforming vector $\fbf_i$ by solving problem (\ref{eq25}). \STATE Compute $d(\mathbf{f}_i ,\mathbf{s}_i;\Hc)$.

\ENDFOR

\STATE Choose  $i^\star = \mathop{\arg\min}_{1\le i\le M}d(\fbf_i,\sbf_i;\Hc).$
\STATE \textbf{Output:} $(\fbf_{i^\star}, \sbf_{i^\star}).$
\end{algorithmic}
\caption{ Greedy Spatial Device Selection (GSDS)}
\label{alg1}
\end{algorithm}

Our  GSDS for device selection is a sequential procedure. We denote the set of selected devices in step $i$  by $\Mc^s_i$, $i=1,2,\ldots,M$.  In the initial step, GSDS selects the device with the strongest channel condition. That is, $\Mc^s_1=\{m_1: m_1=\mathop{\arg\max}_{1\le m\le M}\| \hbf_m\|\}$. In each subsequent step,  a new device is selected based on a metric and added to the set of selected devices. Therefore, $\Mc^s_i$ contains $i$ devices.

   We propose a metric that measures both device channel strength and its correlation to the set of selected devices.
Specifically,
in step $i$, for each unselected device $m \in \Mc\backslash \Mc^s_{i-1}$, we project its channel  $\hbf_m$   onto the subspace spanned by the channel vectors of selected devices.  That is, denote $\Hc^s_{i-1} \triangleq \{\hbf_m: m\in \Mc^s_{i-1}\}$, and let $\text{Proj}_{\Hc^s_{i-1}}(\hbf_m)$ denote the projection of $\hbf_m$ onto $\Hc^s_{i-1}$. GSDS first computes $\|\text{Proj}_{\Hc^s_{i-1}}(\hbf_m)\|$, for each $m \in\Mc\backslash \Mc^s_{i-1}$. It then selects the device, represented by $m_i$, as follows
\begin{align}\label{m_i}
m_i = \mathop{\arg\max}_{m\in \Mc\backslash \Mc^s_{i-1}}\|\text{Proj}_{\Hc^s_{i-1}}(\hbf_m)\|. 
\end{align}
The set of selected devices at step $i$ is then given by 
\begin{align}\label{eq35}
    \Mc^s_i =  \Mc^s_{i-1} \cup \{m_i\}.
\end{align}
 Let $\sbf_i$ be the device selection vector corresponding to $\Mc^s_i$. Once $\sbf_i$ is obtained, we optimize the receiver beamforming vector $\fbf$ to minimize $d(\fbf,\sbf_i;\Hc)$ as in problem \eqref{eq22} with given $\sbf_i$, which is to solve problem \eqref{eq25}, except $\Mc^s$ is replaced by $\Mc^s_i$. Following the approach discussed in Section~\ref{Design}.\ref{GSDS}-\ref{GSDS-sub1}, we obtain the receiver beamforming vector, denoted by $\mathbf{f}_i$ and the value of $d(\mathbf{f}_i , \mathbf{s}_i;\Hc)$. 

After performing all steps $i=1,\ldots,M$, we choose $i^\star$ as
\begin{align}\label{i_star}
i^\star = \mathop{\arg\min}_{1\le i\le M}~d(\fbf_i,\sbf_i;\Hc),
\end{align}
 and the set of selected devices $\Mc^s_{i^\star}$ (and $\sbf_{i^\star}$) and receiver beamforming $\fbf_{i^\star}$ as the output of  GSDS. The detail of GSDS is summarized in Algorithm~\ref{alg1}.

\subsubsection{Computational Complexity}
Algorithm~\ref{alg1} involves obtaining the set of selected devices $\{\Mc^s_i \}_{i=1}^M$ and computing the receiver beamforming vector $\fbf_i$, $i=1,\ldots,M$. For each step $i$, the computational complexity of determining the selected device $m_i$ in \eqref{m_i} is $\mathcal{O}(NM^2)$. Also,    the computational complexity of obtaining the receiver beamforming vector $\fbf_i$ using the SCA method is $\mathcal{O}(I_\text{max} \text{min}(M, N)^3),$ where $I_\text{max} $ is the maximum number of SCA iterations to reach the convergence threshold.  Since there are total $M$ steps in order to determine the set of selected devices $\Mc^s_{i^\star}$ via \eqref{i_star}, the overall computational complexity of GSDS is $\mathcal{O}(I_\text{max} \text{min}(M,N)^3M + NM^3)$.

\subsection{Alternating-optimization-based Device Selection and Beamforming (ADSBF) Approach } \label{ADSBF} The computational complexity of the proposed GSDS grows with the number of devices as $O(M^3)$, which is relatively high as $M$ becomes large. In this section, we propose an algorithm, named Alternating-optimization-based Device Selection and Beamforming (ADSBF), to determine device selection and receiver beamforming with low computational complexity. ADSBF uses an alternating optimization approach to solve problem \eqref{eq22} w.r.t. devices selection $\mathbf{s}$ and receiver beamforming vector $\fbf$ alternatingly. Specifically, ADSBF breaks problem \eqref{eq22} into two subproblems to solve alternatingly: one is the receiver beamforming optimization with given device selection, and the other is device selection under the provided receive beamforming vector. These two subproblems are described below.
\subsubsection{Receiver Beamforming Design Given Device Selection $\mathbf{s}$} \label{ADSBF-sub1}

Given the set of selected devices  $\mathbf{s}$, the minimization of $d(\fbf,\sbf;\Hc)$ in problem \eqref{eq22} w.r.t. $\mathbf{f}$ is given by
    \begin{align}\label{eq25replicate}
    \min_{\mathbf{f}}  & \max_{m \in \Mc^\text{s}}\frac{ K_m^2}{|\mathbf{f}^{H}\mathbf{h}_m|^2}
    \\ &    \textrm{s.t.} \quad   \|\mathbf{f}\| =1.
    \end{align}
which is the same as problem \eqref{eq25}.  Hence, the approach discussed in Section \ref{Design}.\ref{GSDS}-\ref{GSDS-sub1} for transforming the problem into problem \eqref{eq27} is directly applicable to problem \eqref{eq25replicate}.  Following this, we can use the same SCA method to obtain a solution to the problem.

\subsubsection{Device Selection Design Given Receiver Beamforming $\mathbf{f}$}\label{ADSBF-sub2}

Given receiver beamforming vector $\fbf$, problem (\ref{eq22}) reduces to
\begin{align}\label{eq29}
 \min_{\mathbf{s}\in \{ 0,1\}^M} \ &\frac{4}{K^2}(K-\sum_{m=1}^Ms[m]K_m)^2\nn \\
&+\frac{\sigma_n^2}{P_{0}(\sum_{m=1}^M s[m] K_m)^2}\max_{1\le m \le M}\frac{s[m] K_m^2}{|\mathbf{f}^{H}\mathbf{h}_m|^2}.
\end{align}
The above problem is an integer program, and it also contains a min-max optimization problem, which typically is hard to solve. However, for this problem, we are able to develop an efficient algorithm to solve it.  

Specifically, we first sort  $\big\{\frac{K_m^2}{|\fbf^H \hbf_m|^2}\big\}$ in ascending order and index the corresponding devices as $m_1,\cdots,m_M$:
\begin{align}
\frac{K_{m_1}^2}{|\mathbf{f}^{H}\mathbf{h}_{m_1}|^2}\le \frac{K_{m_2}^2}{|\mathbf{f}^{H}\mathbf{h}_{m_2}|^2} \le ...\le \frac{K_{m_M}^2}{|\mathbf{f}^{H}\mathbf{h}_{m_M}|^2}.
\end{align}
Let $c\triangleq \max_{1\le m \le M}\frac{s[m] K_m^2}{|\mathbf{f}^{H}\mathbf{h}_m|^2}$, which is attained by some $m_j$, for $1\le j\le M$.   This means that  $s[m_{j'}]=0$, $j'>j$, and only the first $j$ sorted devices, $m_1,\ldots,m_j$, are candidates for selection, and the rest are not selected. Next, for fixed $c$, note that the objective function in \eqref{eq29} decreases as more devices  from $\{m_1, \ldots,m_j\}$ are selected. As a result, the minimum objective value is attained by selecting all these devices, $m_1,\ldots,m_j$. 

Therefore, let $\mathbf{z}_j$, $1\le j\le M$, denote the device selection vector  that selects  devices $m_1, \ldots,m_j$, given by 
\begin{align}
\mathbf{z}_j[m]=  \begin{cases} 1, & \text{for } m \in \{m_1, \ldots,m_j\}\\0,&\text{otherwise}.\end{cases}
\end{align}
We  evaluate the objective function value $d(\mathbf{f}, \mathbf{z}_j;\Hc)$ under each device selection vector   $\zbf_j$, for  $j=1,\ldots,M$. Then, we obtain the optimal selection vector $\mathbf{s}$ from $\{ \mathbf{z}_j\}$ that achieves the minimum value among all  $d(\mathbf{f}, \mathbf{z}_j;\Hc)$'s: \begin{align}
j^\star &= \underset{1\le j\le M}{\text{argmin}}\:d(\mathbf{f}, \mathbf{z}_j;\Hc), \\
\mathbf{s} &= \mathbf{z}_{j^\star}.
\end{align}

We summarize the device selection algorithm in Algorithm \ref{alg2}. Below, we show that our proposed algorithm is guaranteed to find the optimal solution $\mathbf{s}$ to problem \eqref{eq29}.

\begin{algorithm}[t]
\begin{algorithmic}[1]
\STATE \textbf{Input:} $\mathbf{f}$, $\{\mathbf{h}_m, K_m: m \in \Mc\}$.
\STATE Sort $\big\{\frac{K_m^2}{|\mathbf{f}^{H}\mathbf{h}_m|^2}\big\}$ in ascending order: 
\begin{align*}
\frac{K_{m_1}^2}{|\mathbf{f}^{H}\mathbf{h}_{m_1}|^2}\le \frac{K_{m_2}^2}{|\mathbf{f}^{H}\mathbf{h}_{m_2}|^2} \le ...\le \frac{K_{m_M}^2}{|\mathbf{f}^{H}\mathbf{h}_{m_M}|^2},
\end{align*}
with  device indices $m_1,\ldots,m_M$. 
\FOR {$j=1,\ldots,M$}
\STATE Set
\begin{align*}
\mathbf{z}_j[m]=  \begin{cases} 1, & \text{for } m \in \{m_1, \ldots,m_j\}\\0,&\text{otherwise}.\end{cases}
\end{align*}
\ENDFOR
\STATE Choose  $j^\star= \underset{1\le j\le M}{\text{argmin}}\:d(\mathbf{f}, \mathbf{z}_j;\Hc)$ and set $\mathbf{s}^{\star}=\mathbf{z}_{j^\star}$.
\STATE \textbf{Output:} $\sbf^\star$
\end{algorithmic}
\caption{ Optimal Device Selection Given Beamforming}
\label{alg2}
\end{algorithm}
 
\begin{proposition}
 Algorithm~\ref{alg2} produces  a global optimal point for problem (\ref{eq29}).
\end{proposition}

\IEEEproof Assume $\mathbf{y}$ is an arbitrary device selection vector. Let $m^\dagger$ be the device with the largest value of $\frac{K_m^2}{|\mathbf{f}^{H}\mathbf{h}_m|^2}$ among the selected devices in $\mathbf{y}$. Assume its corresponding index in  the sorted devices $\{m_1,\ldots,m_M\}$ in Algorithm~\ref{alg2}  is $m_{j^\dagger}$ (\ie $m^\dagger=m_{j^\dagger}$). Then, the set of selected devices in $\mathbf{y}$ is  a subset of $\{m_1, m_2,..., m_{j^\dagger}\}$, \ie the devices selected in $\zbf_{j^\dagger}$ defined by Algorithm 2.
From the objective function in \eqref{eq29}, we have $d(\mathbf{f}, \mathbf{z}_{j^\dagger};\Hc) \le d(\mathbf{f}, \mathbf{y};\Hc) $. Thus, for $\forall\ybf\in\{0,1\}^M$, we can find a  selection vector in $\{\zbf_j\}$ with an equal or less objective value. Thus,   the global optimal point is in $\{\zbf_j\}$.
\endIEEEproof

Our proposed ADSBF alternatingly solve the two subproblems w.r.t.  $\mathbf{f}$ and $\mathbf{s}$ described in Sections~\ref{Design}.\ref{ADSBF}.\ref{ADSBF-sub1} and \ref{Design}.\ref{ADSBF}.\ref{ADSBF-sub2}, respectively, to obtain a solution to problem \eqref{eq22}. The overall ADSBF algorithm is summarized in Algorithm \ref{alg3}.

\subsubsection{Initialization and Convergence}
Note that the SCA method used for the receiver beamforming subproblem requires a feasible initial point to problem \eqref{eq27}. At the initial iteration $l=1$, this initial point can be found by solving problem \eqref{eq27} using the semi-definite relaxation (SDR) approach \cite{dong2020multi}, which gives a feasible approximate solution. In the subsequent iteration  $l > 1 $, the receive beamforming vector $\fbf^{(l-1)}$ obtained from the previous iteration $(l-1)$ is used as the initial point for the SCA method in this iteration $l$. Since the SCA is guaranteed to converge to a local minimum, and the device selection subproblem is solved optimally, the objective value in \eqref{eq22} is non-increasing over iterations and is non-negative. Thus, ADSBF is guaranteed to converge.

\begin{algorithm}[t]
\begin{algorithmic}[1]
\STATE \textbf{Initialization:} Set $\epsilon$. Set initial $\fbf^{(0)}$ with $\|\fbf^{(0)}\|=1$, and $\mathbf{s}^{(0)}={\bf 1}$. Set $l=0$.
\REPEAT
   \STATE Obtain $\mathbf{f}^{(l+1)}= \underset{ \{\mathbf{f}: \| \mathbf{f} \|=1 \}}{\text{argmin}}\: d(\mathbf{f} , \mathbf{s}^{(l)};\Hc)$ by the SCA method, where  $\mathbf{f}^{(l)}$ is used as the initial point for SCA.
   \STATE Obtain $\mathbf{s}^{(l+1)}\!=\! \underset{ \mathbf{s} \in \{0,1\}^M}{\text{argmin}} d(\mathbf{f}^{(l+1)} , \mathbf{s;\Hc})$ by {Algorithm \ref{alg2}}.   
\STATE Set $l\leftarrow l+1$.    
\UNTIL $|d(\mathbf{f}^{(l)} , \mathbf{s}^{(l)};\Hc) - d(\mathbf{f}^{(l-1)} , \mathbf{s}^{(l-1)};\Hc)| \le \epsilon$ 
\STATE Set $\fbf^\star = \fbf^{(l)}$, $\sbf^\star = \sbf^{(l)}$.
\STATE \textbf{Output:} $\fbf^{\star}$, $\sbf^{\star}$.
\end{algorithmic}
\caption{Alternating-Optimization-Based Device Selection and Beamforming (ADSBF)}
\label{alg3}
\end{algorithm}
\subsubsection{Computational Complexity}
The computational complexity of Algorithm \ref{alg2} is $\mathcal{O}(M\text{log}(M)+MN)$, and that for solving problem (\ref{eq27}) by the SCA method is $\mathcal{O}(I_{\text{max}} \min(N,M)^3)$, where $I_{\text{max}}$ is the  number of SCA iterations. Therefore, the overall computational complexity of ADSBF is $\Oc(J_{\text{max}}(I_{\text{max}} \min(N,M)^3+ M\text{log}(M)+MN)),$ where $J_{\text{max}}$ is the maximum number of iterations of ADSBF. 

In comparison, the computational complexity of the Gibbs sampling approach proposed by \cite{liu2021reconfigurable} is $\Oc(M^4)$.
Also, the computational complexity of the DC approach in \cite{yang2020federated} is $\Oc(M(M+N^2)^3+ M N^6)$. We summarize in Table \ref{ComplexityTable} the computational complexities of different methods when  $M \gg N$.  We see the computational advantage of GSDS and ADSBF over the existing approaches. In particular, ADSBF has substantially lower complexity than all other approaches.

\begin{table}[h]
\caption{Computational complexity comparison when $M\gg N$.}
\label{ComplexityTable}
\setlength{\tabcolsep}{3pt}
\centering
\begin{tabular}{|p{50pt}|p{30pt}|p{70pt}|p{30pt}|}
\hline
\textbf{ADSBF} & \textbf{GSDS} & \textbf{Gibbs Sampling\cite{liu2021reconfigurable}} & \textbf{DC\cite{yang2020federated}}\\
\hline
$\Oc(M\text{log}(M))$& $\mathcal{O}(M^3)$ &   $\Oc(M^4)$& $\Oc(M^4)$ \\
\hline
\end{tabular}
\end{table}

\section{Simulation Results}

In this section, we evaluate the performance of our proposed approaches for an image classification task. We consider training and testing using the logistic regression for the MNIST \cite{MNIST} dataset and the convolutional neural networks (CNNs) for the CIFAR-$10$ \cite{CIFAR10} dataset.\footnote{The simulation codes are available online at \url{https://github.com/faezemoradik/BeamformingDeviceSelecetionDesignInFLwithOTA}.}

We consider a scenario with $M=200$ devices and $N=16$ antennas. The distance between each device $m$ and the parameter server, denoted by $d_m$, is drawn from a uniform distribution: $d_m \sim U[r_{\text{min}}, r_{\text{max}}] $, where $r_{\text{min}} = 10~\text{m}$ and $r_{\text{max}} = 100~\text{m}$. 
The path loss follows the COST Hata model, given by $\text{PL[dB]} = 139.1 + 35.22 \: \text{log}(d_m[\text{km}])$. We assume the device channels are constant during the training. The channel vector for device $m$ is generated using a complex Gaussian distribution as $\mathbf{h}_{m,t}= \mathbf{h}_m \sim \mathcal{CN}(\mathbf{0}, \frac{1}{PL}\mathbf{I}_{N}), \forall t$. The maximum permissible average transmit power for the devices is assumed to be $P_{0} = 0$ dBm. For  comparison, we also consider  the following  approaches:
\begin{enumerate}
    \item \textbf{Select all:} All of the devices are selected to contribute to the FL training, i.e., $\mathcal{M}^s= \mathcal{M}$.\ The receive beamforming is obtained by solving problem \eqref{eq25} via the SCA method.
    
    \item \textbf{Top one:} Only the device with the strongest channel condition is selected to contribute to the FL training, i.e., $\mathcal{M}^s= \{m^{\dagger}\}$, where $m^{\dagger}= \arg\min_{1\le m\le M}\: \|\mathbf{h}_m\|$.   The receiver beamforming vector is aligned with the channel vector of the selected device, i.e., $\mathbf{f}= {\mathbf{h}_{m^\dagger}}/{\| \mathbf{h}_{m^\dagger}\|}$.
    
    \item \textbf{Gibbs sampling \cite{liu2021reconfigurable}:} Gibbs sampling is used for device selection, and the receiver beamforming vector is obtained by an SCA method.
    
    \item \textbf{DC approach \cite{yang2020federated}:} The receiver beamforming and the device selection are jointly optimized by DC programming to maximize the number of selected devices, subject to the  MSE of the over-the-air aggregation for the global model is no larger than threshold $\gamma$.   
\end{enumerate}

Note that for a fair comparison of the computational complexity, the SCA method described in Section \ref{Design}.\ref{GSDS}.\ref{GSDS-sub1} is used in GSDS, ADSBF, ``Select all", and Gibbs sampling methods. The maximum number of iterations for ADSBF is set to $J_{\text{max}} = 10$. For the Gibbs sampling method, to achieve the optimal performance, we set the initial temperature $\beta_0=1$, and the cooling schedule parameter $\rho= 0.1$. Additionally, the number of iterations is configured to be $40$. It is important to note that these parameter values have been determined through hyper-parameter tuning. For DC approach, the parameter $\gamma$ must be carefully tuned to ensure that an appropriate number of devices are selected to attain the optimal performance. After conducting the hyper-parameter tuning, we set $\gamma = 94$ dB to achieve the fastest training convergence. Note that since the channel conditions in our experiments are much weaker than that of \cite{yang2020federated}, the resulting MSE is larger, and hence we need to set $\gamma$ to a larger value compared to \cite{yang2020federated}.
In contrast to Gibbs sampling and DC approach, our proposed GSDS and ADSBF do not require extra hyperparameter tuning. This is a noticeable advantage, as our proposed methods do not need extensive tuning.

\subsection{MNIST Dataset}

In the MNIST dataset, each individual data sample is a labeled gray-scale image of $28 \times 28$ pixels, depicting a handwritten digit, denoted by $\mathbf{x}_k \in \mathbb{R}^{784}$. The corresponding label, $y_k \in \{0, 1, ..., 9\}$, specifies the class to which the image belongs. The dataset consists of $60,000$ training samples and $10,000$ test samples, belonging to ten different classes. We consider training a multinomial logistic regression classifier with the cross-entropy loss function given by
    \begin{align}\label{eq36}
    l(\mathbf{w}; \mathbf{x}_k , y_k) = - \sum \limits_{j=0}^9 1\{y_k = j\} \text{log}\frac{\text{exp}(\mathbf{u}_k^T\mathbf{w}^{(j)})}{\sum \limits_{i=0}^9 \text{exp}(\mathbf{u}_k^T\mathbf{w}^{(i)})}, 
    \end{align}
where $1\{\cdot\}$ is an indicator function, $\ubf_k =[\mathbf{x}_k^T,1 ]^T$, and $\mathbf{w}= [{\mathbf{w}^{(0)}}^T,\ldots,{\mathbf{w}^{(9)}}^T]^T$ with $\wbf^
{(j)}\in \mathbb{R}^{785}$ being the model parameter vector for class $j$, consisting of 784 weights and a bias term. 

We assume the data distribution over devices is i.i.d. Each device's local dataset contains an equal number of data samples from different classes, and the number of local data samples in each device is $K_m =270$. Upon thorough hyperparameter tuning, we set a unified learning rate $\lambda = 0.05$  for the global model update in \eqref{eq5} for all approaches. Furthermore, the full local batch is used for the gradient computation in each communication round for all approaches. A (training) epoch refers to one training pass over the entire training dataset by all devices. Since the devices are using the full local batch in this setting, each epoch only consists of a single communication round.
Finally, we set the receiver noise power  $\sigma_n^2 =-20$ dBm, which generally includes both receiver thermal noise and interference signals. 

\begin{figure}[t]
\centerline{\includegraphics[width=0.5\textwidth]{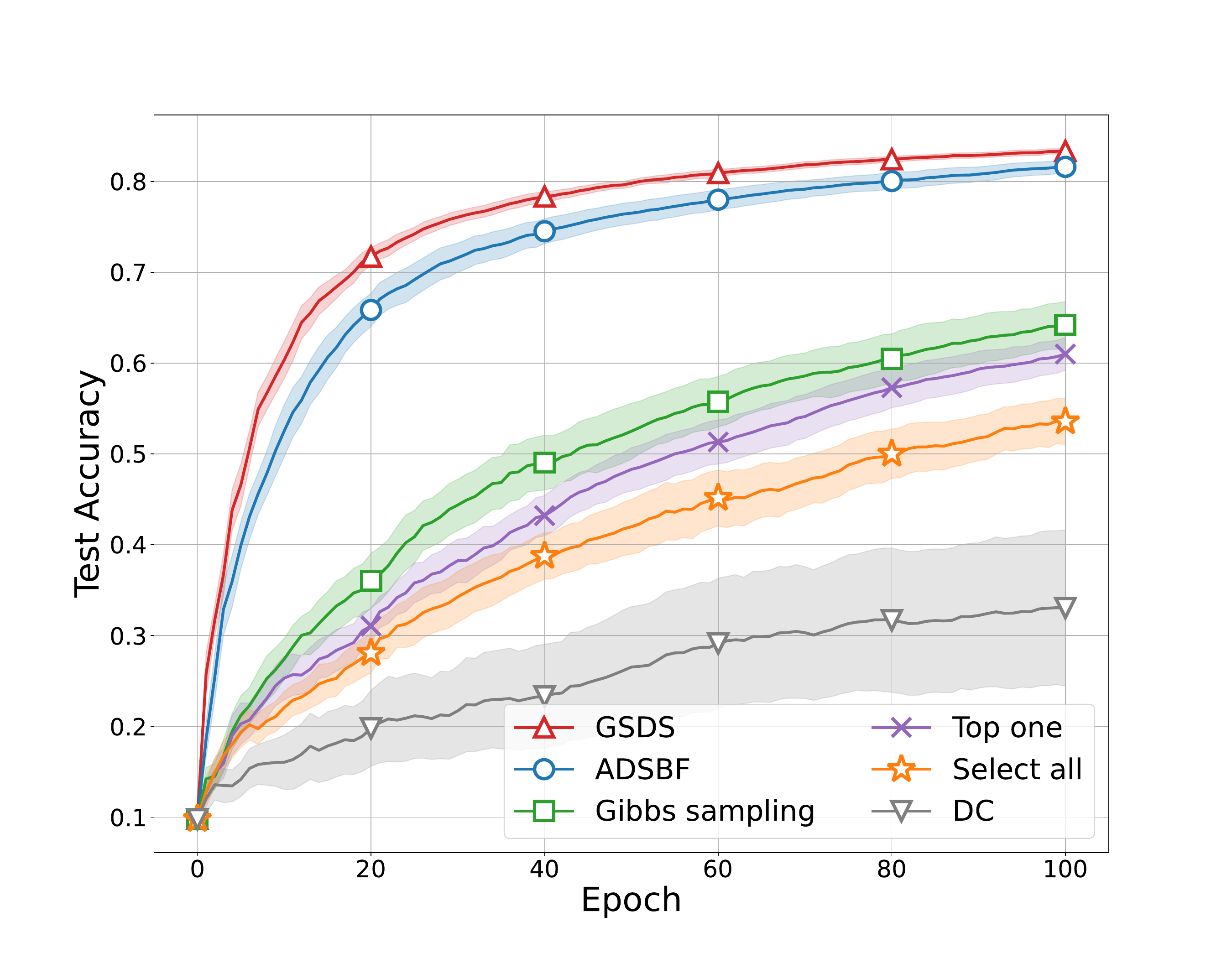}}
\caption{ Average test accuracy over epoch (MNIST dataset).}
\label{acc_mnist}
\centerline{\includegraphics[width=0.5\textwidth]{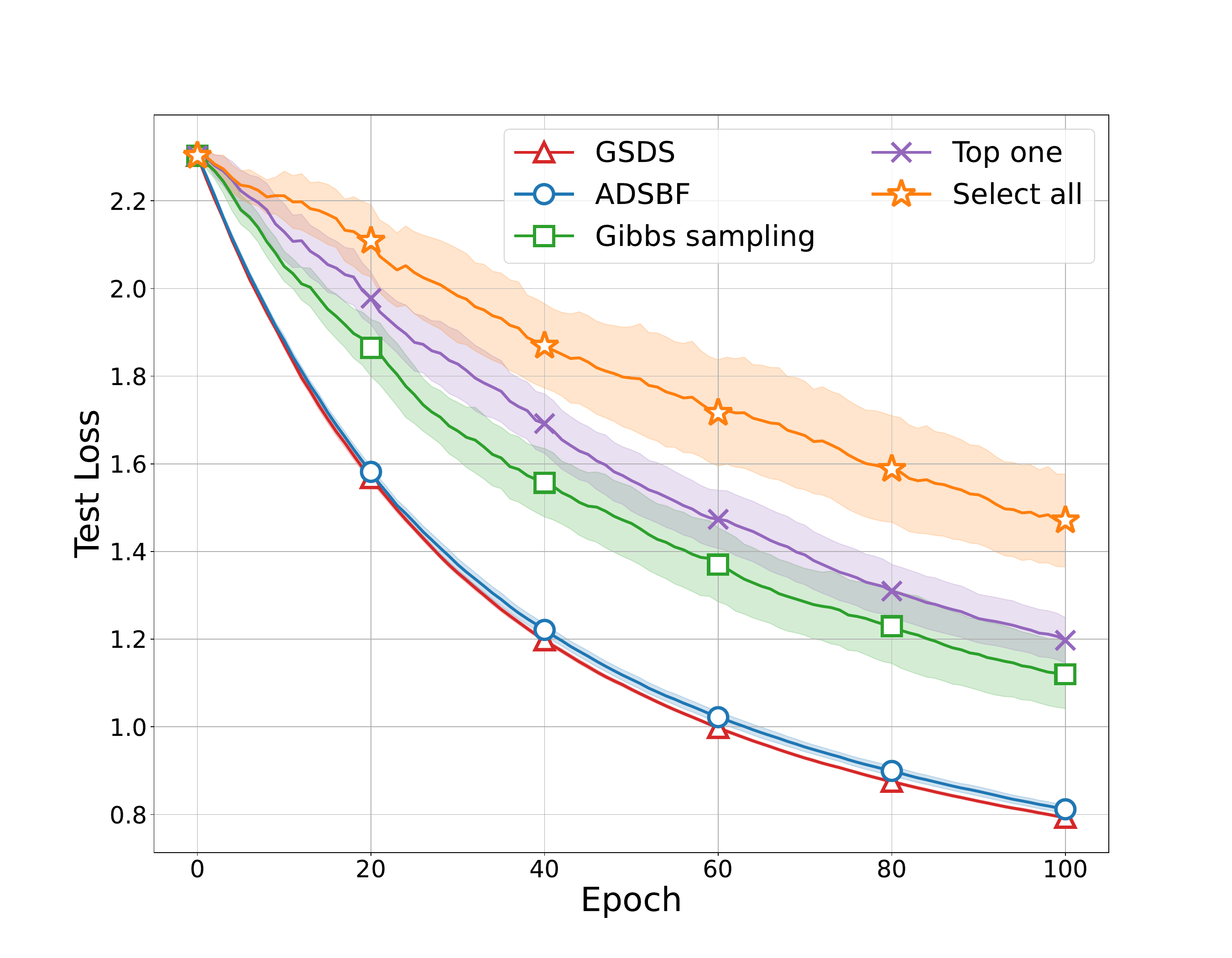}}
\caption{Average global test loss over epoch (MNIST dataset).}
\label{loss_mnist}
\end{figure}
\begin{figure}
\centerline{\includegraphics[width=0.5\textwidth]{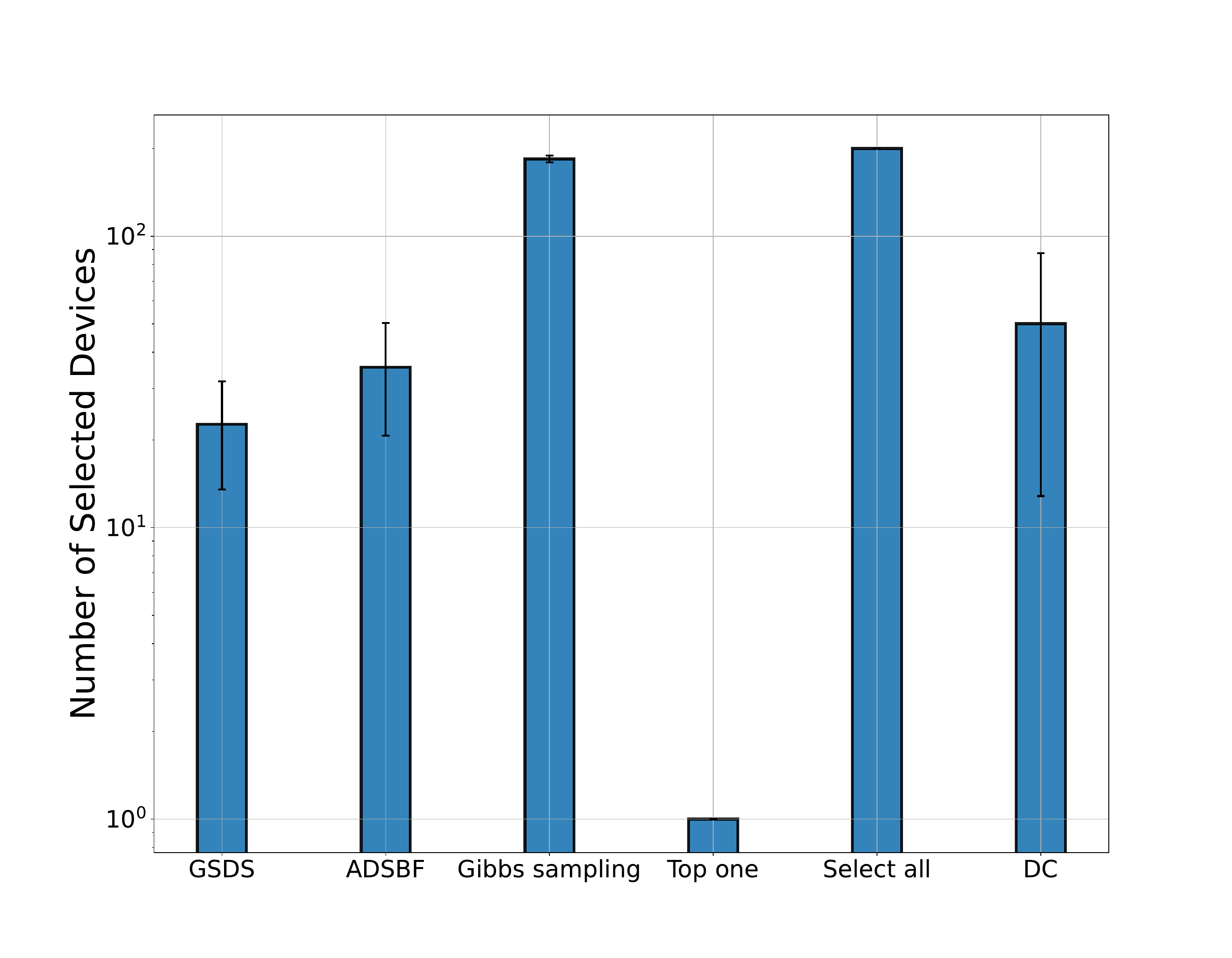}}
\caption{Average number of selected devices for different approaches (MNIST dataset).}
\label{num_selected_mnist}
\end{figure}

Figs.~\ref{acc_mnist} and~\ref{loss_mnist} respectively show the average test accuracy and the average test loss with $95\% $ confidence intervals over the $20$ different channel realizations and receiver noise realizations. 
We see from both figures that our proposed GSDS and ADSBF approaches converge after $T=100$ rounds, which is the fastest convergence rate among all the considered methods. Also,  they achieve the highest test accuracy and lowest test loss among the approaches considered. 
In particular,  GSDS and ADSBF provide test accuracy above $80\%$, which is $15\%$ higher than the best accuracy achieved by the benchmarks. This significant improvement demonstrates the efficacy of our proposed methods for device selection. Between ADSBF and GSDS, although ADSBF performs slightly worse than GSDS, its run time is significantly lower than that of GSDS as discussed below.
Note that due to high instability, the loss of DC is significantly larger than that of the other approaches, so we omit it in Fig.~\ref{loss_mnist}.

Fig.~\ref{num_selected_mnist} shows the average number of selected devices by different approaches over $20$  channel realizations. The accompanying error bar indicates the standard deviation around the average value. We see that ADSBF and GSDS select fewer devices than Gibbs sampling and DC approaches, but more than ``Top one". This result, combined with the learning performance results in Figs.~\ref{acc_mnist} and~\ref{loss_mnist}, demonstrates the effectiveness of our methods in providing a proper trade-off between imperfect noisy communication and the amount of training data provided by the local devices for model training through device selection.  

Table~\ref{tab2} lists the average computation time of generating the beamforming and device selection solution using different approaches prior to the start of training. As we see in the column of the MNIST dataset, our proposed  GSDS and ADSBF have significantly lower computational complexity as compared with the Gibbs sampling and DC approach.
Furthermore, we see that the run time of ADSBF is two magnitudes lower than that of GSDS, with a slightly worse test accuracy. This demonstrates the computational advantage of ADSBF over GSDS and the overall efficacy of ADSBF.

\begin{table}[t]
\caption{Average run time for different approaches (seconds)}
\label{table}
\setlength{\tabcolsep}{3pt}
\begin{tabular}{|p{60pt}||p{70pt}| p{70pt} |}
\hline
\textbf{Method} & 
\textbf{MNIST Dataset } & \textbf{CIFAR-10 Dataset} \\
\hline
GSDS & 585.7 & 664.5\\
\hline
ADSBF & 8.1 & 9.3\\
\hline
Gibbs sampling & 43410.5 &49799.3 \\
\hline
DC & 2875.9 & 2858.7\\
\hline
Select all & 6.2 & 7.2 \\
\hline
Top one & 0.002 & 0.007\\
\hline
\end{tabular}
\label{tab2}
\end{table}

\subsection{CIFAR-10 Dataset}

For the CIFAR-10 dataset, each individual data sample is a colored image of $3 \times 32 \times 32$ pixels, which is represented as $\mathbf{x}_k \in \mathbb{R}^3 \times \mathbb{R}^{32} \times \mathbb{R}^{32}$; the associated label $y_k \in \{0, 1, ..., 9\}$ indicates the image's corresponding class. The dataset comprises a total of $50,000$ training samples and $10,000$ test samples. Given the increased complexity of this dataset, we opt for a more sophisticated convolutional neural network (CNN) model: the Residual Network (ResNet) with 14 layers (ResNet-14) \cite{ResNet}. The training process employs the cross-entropy loss function. To enhance the training process, we implement a data augmentation technique outlined in \cite{ResNet}. This technique involves augmenting the data by adding a 4-pixel padding on all sides and randomly selecting a $32 \times 32$ crop from either the padded image or its horizontally flipped counterpart.

Note that our joint optimization problem \eqref{eq22} is obtained based on the training convergence analysis under Assumptions  A1-A4 with a strongly convex loss function, which is not the case here due to the non-convex nature of CNNs. Nonetheless, we test our proposed approaches using this dataset to demonstrate the effectiveness of our proposed method for this application.

Following the approach similar to that for the MNIST dataset, we uniformly allocate the training samples of each class across the devices, with a total number of samples per device as $K_m = 250$.
During each communication round, the devices compute their local gradient by processing a batch of 50 data samples from their respective local datasets. Consequently, each epoch comprises 5 communication rounds, and we assess the model's accuracy upon the completion of each epoch. We use the built-in SGD optimizer in PyTorch \cite{NEURIPS2019_9015} and set the value of the learning rate as $0.01$, the value of momentum as  $0.9$, and the value of weight decay factor as $10^{-4}$. 
We set the receiver noise power $\sigma_n^2=-50$ dBm.

\begin{figure}[t]
\centerline{\includegraphics[width=0.5\textwidth]{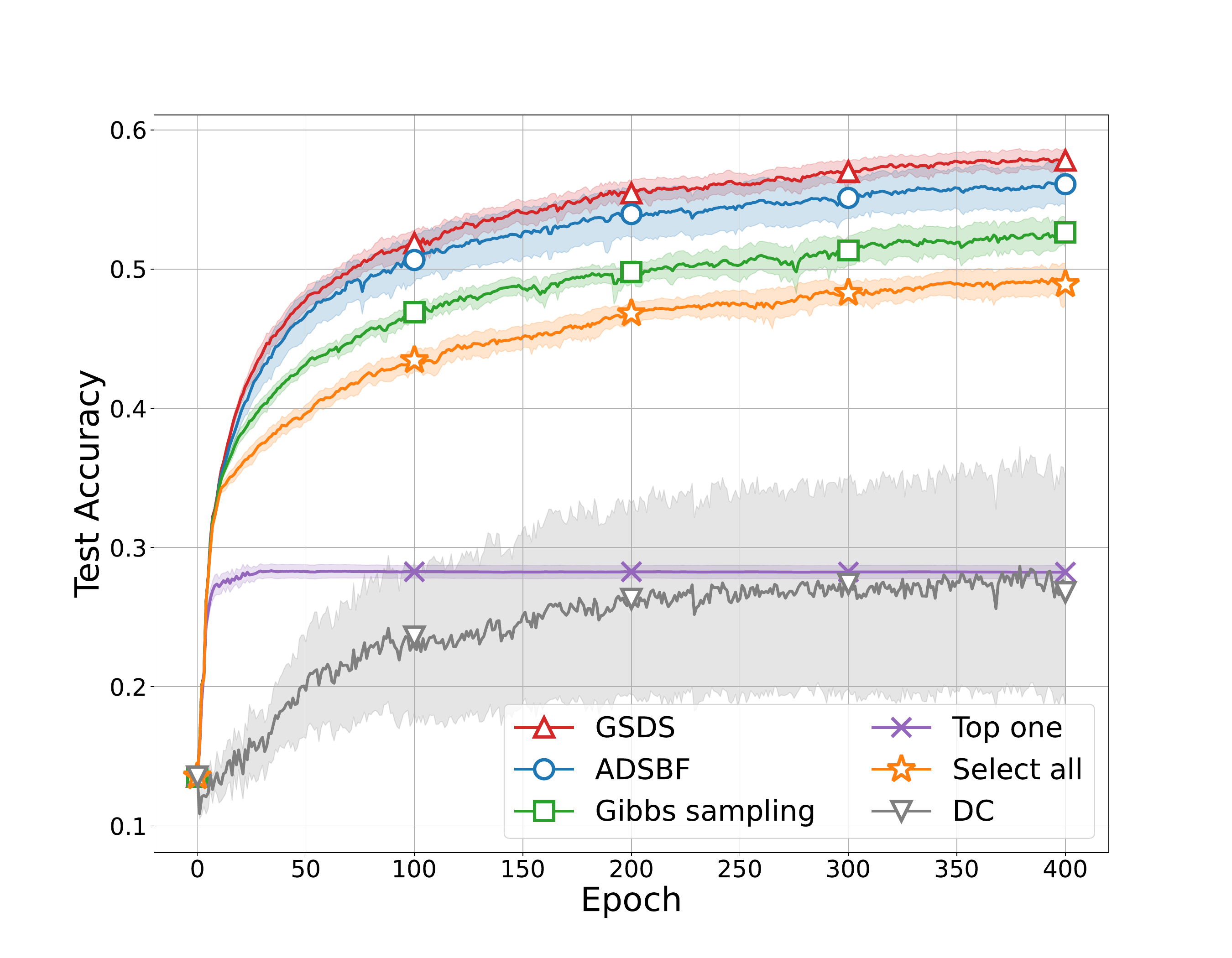}}
\caption{ Average test accuracy over epoch (CIFAR-10 dataset).}
\label{acc_cifar}
\centerline{\includegraphics[width=0.5\textwidth]{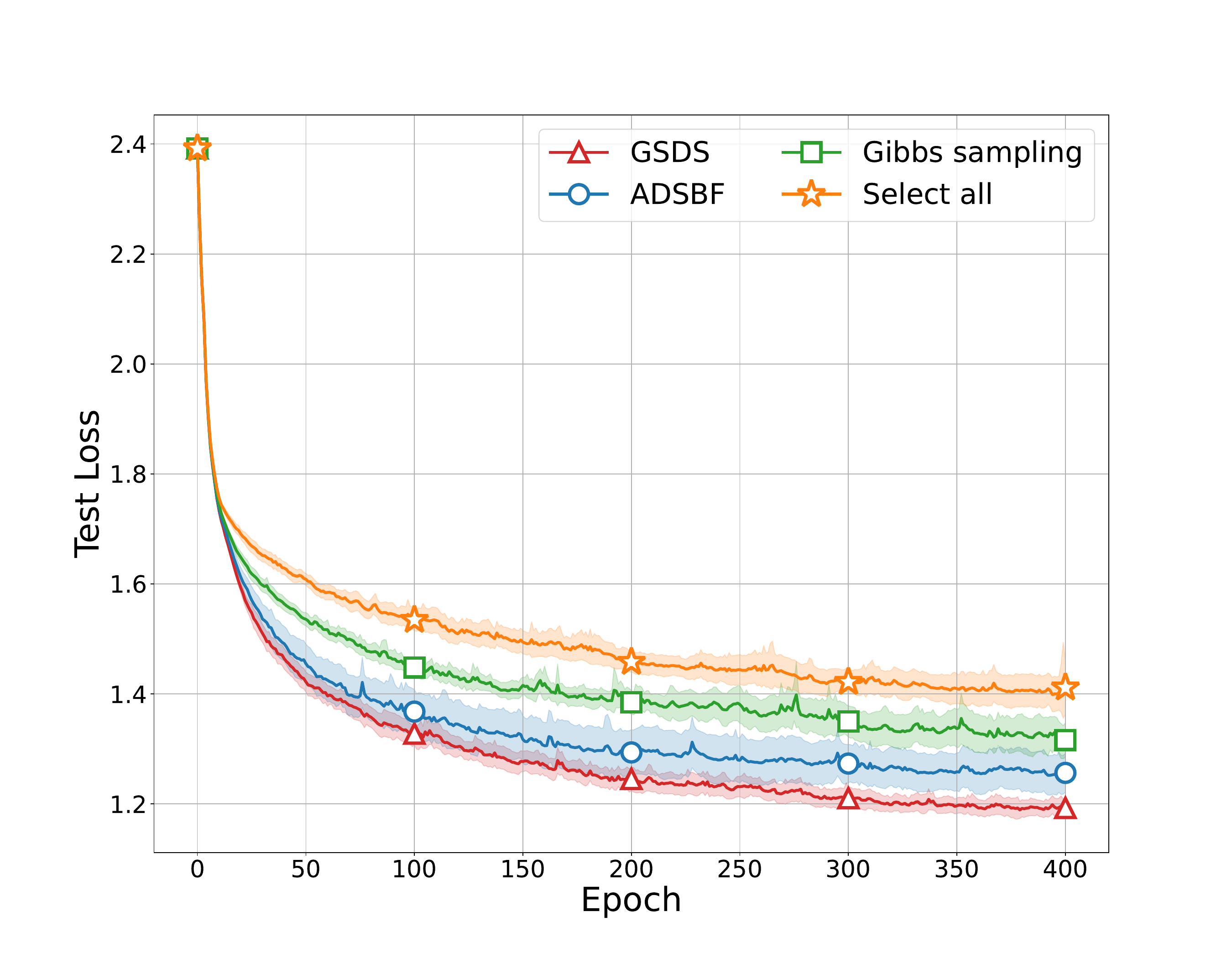}}
\caption{Average global test loss over epoch (CIFAR-10 dataset).}
\label{loss_cifar}
\end{figure}
\begin{figure}
\centerline{\includegraphics[width=0.5\textwidth]{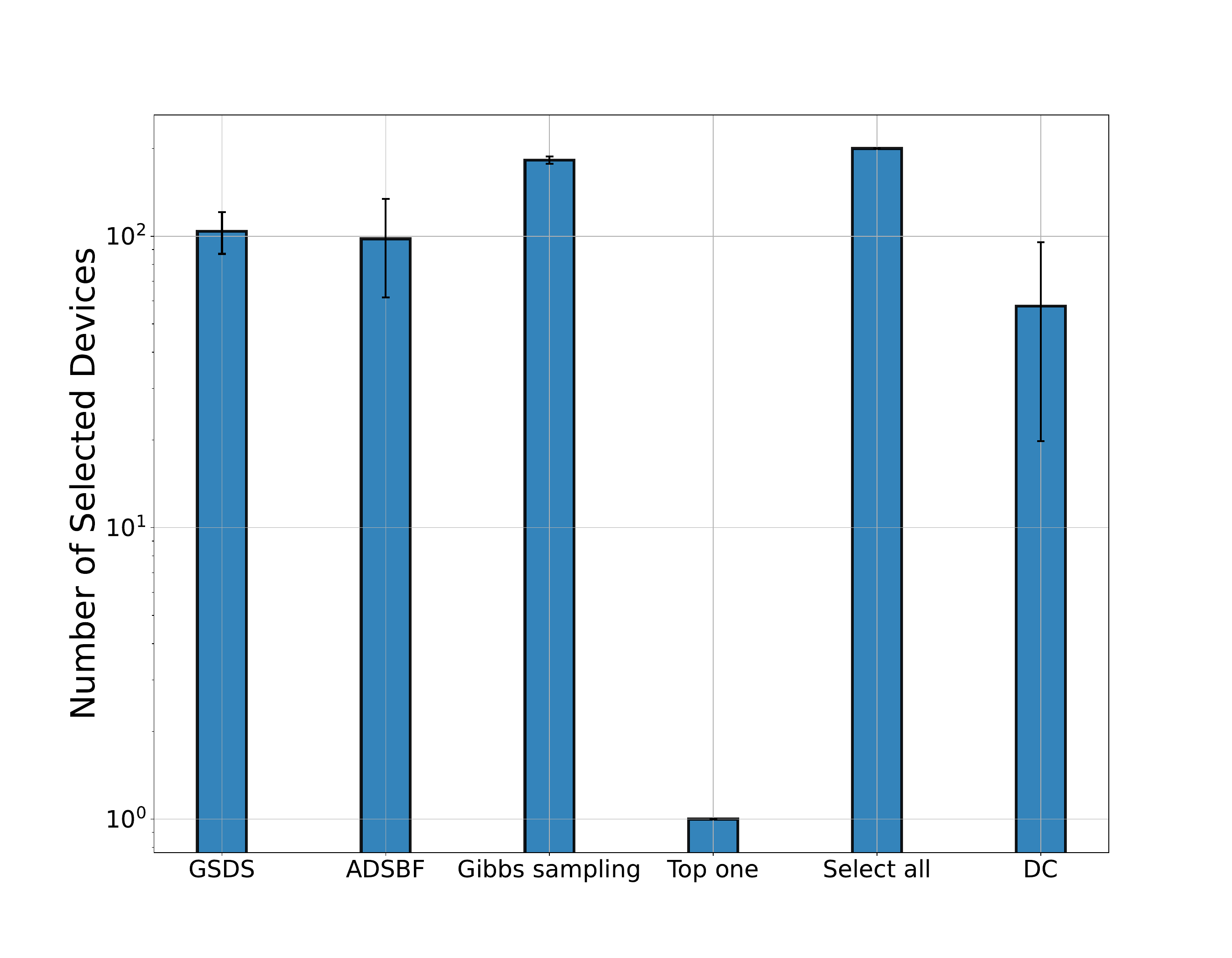}}
\caption{Average number of selected devices for different approaches (CIFAR-10 dataset).}
\label{num_selected_cifar}
\end{figure}

Figs.~\ref{acc_cifar} and ~\ref{loss_cifar} show the average test accuracy and average test loss, along with $95\%$ confidence intervals, across $20$ different channel and communication noise realizations.
Again, we see that our proposed GSDS and ADSBF  consistently achieve superior test accuracy and lower test loss after $400$ epochs as compared to other approaches. Specifically, the final test accuracy achieved by GSDS is approximately $58\%$, while the highest test accuracy among the benchmarks is
$53\%$.
Note that we omit showing the loss values for ``Top one" and DC in Fig. \ref{loss_cifar} as these methods result in a very large test loss due to their low accuracy, as shown in Fig. \ref{acc_cifar}. 

Comparing Figs.~\ref{acc_mnist} and ~\ref{acc_cifar} using two different datasets, we note that ``Top one" performs better than ``Select all" on the MNIST dataset, while ``Select all" performs better than ``Top one" on the CIFAR-10 dataset. The reason for this observation is as follows: The CIFAR-10 dataset is more difficult to learn compared with the MNIST dataset, as the variation among different data samples is much higher in CIFAR-10  compared with that in MNIST. Therefore, for the MNIST dataset, even if we only use the data samples stored in a single device (\ie  ``Top one") to train the model, the achieved test accuracy can still be relatively high. On the other hand, ``Select all"  suffers from a relatively large aggregation error caused by devices having weak channels, resulting in worse performance.
However, for the CIFAR-10 dataset,  if we train the model exclusively with the samples from a single device, the performance will suffer as the variation among the samples cannot be adequately captured during training.

The beamforming and device selection computation time for each approach, prior to the start of training for the ResNet model, is listed in the last column of Table \ref{tab2}. Similar to that of the MNIST dataset, our proposed methods have significantly lower computational complexity than the Gibbs sampling and DC approaches.

Fig. \ref{num_selected_cifar} shows the average number of selected devices and the standard deviation over 20 channel realizations obtained by different approaches. Similar to the experiments with the MNIST dataset, we see that ADSBF and GSDS select fewer devices than Gibbs sampling, but more than DC and ``Top one". This again shows the effectiveness of our methods in choosing an appropriate set of devices to trade off communication imperfection and the amount of training data from devices to achieve a satisfactory learning performance.

\section{Conclusion}
In this paper, we have jointly designed uplink receiver beamforming and device selection in over-the-air FL to minimize the global training loss after arbitrary $T$ communication rounds, assuming time-varying wireless channels. To tackle this challenging stochastic optimization problem, we have obtained an upper bound for the global training loss and designed receiver beamforming and device selection to minimize this upper bound.
We have proposed two approaches, GSDS and ADSBF, to obtain a solution.
GSDS  uses a greedy method that exploits the channel strength and correlation to sequentially add devices to the set of selected devices. In contrast, ADSBF employs the alternating optimization technique to solve the device selection and receiver beamforming subproblems alternatingly, where we have provided an efficient algorithm to solve the device selection subproblem optimally with low computational complexity. In both approaches, we have shown that given the selected devices, the receiver beamforming optimization problem is equivalent to downlink single-group
multicast beamforming, for which existing efficient algorithms can be used to obtain a solution. The simulation results obtained from image classification experiments have demonstrated that both GSDS and ADSBF speed up the training convergence and have lower computational complexity than the state-of-the-art approaches. Furthermore, we have observed that GSDS and ADSBF offer two distinct design choices with a trade-off between learning performance and computational complexity.

\bibliographystyle{IEEEbib}
\bibliography{Refs}

\begin{thebibliography}{10}

\bibitem{zhu2020toward}
G. Zhu, D. Liu, Y. Du, C. You, J. Zhang, and K. Huang,
\newblock ``Toward an intelligent edge: Wireless communication meets machine learning,''
\newblock {\em {IEEE} Commun. Mag.}, vol. 58, no. 1, pp. 19--25, 2020.

\bibitem{lim2020federated}
W.~Y.~B. Lim, N.~C. Luong, D.~T. Hoang, Y. Jiao, Y.-C. Liang, Q. Yang, D. Niyato, and C. Miao,
\newblock ``Federated learning in mobile edge networks: A comprehensive survey,''
\newblock {\em {IEEE} Commun. Surveys Tuts}, vol. 22, no. 3, pp. 2031--2063, 2020.

\bibitem{zhu2019broadband}
G. Zhu, Y. Wang, and K. Huang,
\newblock ``Broadband analog aggregation for low-latency federated edge learning,''
\newblock {\em {IEEE} Trans. Wireless Commun.}, vol. 19, no. 1, pp. 491--506, 2019.

\bibitem{amiri2020machine}
M.~M. Amiri and D. G{\"u}nd{\"u}z,
\newblock ``Machine learning at the wireless edge: Distributed stochastic gradient descent over-the-air,''
\newblock {\em {IEEE} Trans. Signal Process.}, vol. 68, pp. 2155--2169, 2020.

\bibitem{amiri2020federated}
M.~M. Amiri and D. G{\"u}nd{\"u}z,
\newblock ``Federated learning over wireless fading channels,''
\newblock {\em {IEEE} Trans. Wireless Commun.}, vol. 19, no. 5, pp. 3546--3557, 2020.

\bibitem{sery2020analog}
T. Sery and K. Cohen,
\newblock ``On analog gradient descent learning over multiple access fading channels,''
\newblock {\em {IEEE} Trans. Signal Process.}, vol. 68, pp. 2897--2911, 2020.

\bibitem{guo2020analog}
H. Guo, A. Liu, and V.~K. Lau,
\newblock ``Analog gradient aggregation for federated learning over wireless networks: Customized design and convergence analysis,''
\newblock {\em IEEE Internet of Things Journal}, vol. 8, no. 1, pp. 197--210, 2020.

\bibitem{zhang2021gradient}
N. Zhang and M. Tao,
\newblock ``Gradient statistics aware power control for over-the-air federated learning,''
\newblock {\em {IEEE} Trans. Wireless Commun.}, vol. 20, no. 8, pp. 5115--5128, 2021.

\bibitem{zhang2021federated}
J. Zhang, N. Li, and M. Dedeoglu,
\newblock ``Federated learning over wireless networks: A band-limited coordinated descent approach,''
\newblock in {\em Proc. {IEEE} Conf. on Computer Communications (INFOCOM)}, 2021.

\bibitem{wang2022online}
J. Wang, M. Dong, B. Liang, G. Boudreau, and H. Abou-zeid,
\newblock ``Online model updating with analog aggregation in wireless edge learning,''
\newblock in {\em Proc. {IEEE} Conf. on Computer Communications (INFOCOM)}, 2022, pp. 1229--1238.

\bibitem{PracticalImplementation1}
H. Guo, Y. Zhu, H. Ma, V.~K.~N. Lau, K. Huang, X. Li, H. Nong, and M. Zhou,
\newblock ``Over-the-air aggregation for federated learning: Waveform superposition and prototype validation,''
\newblock {\em Journal of Communications and Information Networks}, vol. 6, no. 4, pp. 429--442, 2021.

\bibitem{PracticalImplementation2}
L. Li, C. Huang, D. Shi, H. Wang, X. Zhou, M. Shu, and M. Pan,
\newblock ``Energy and spectrum efficient federated learning via high-precision over-the-air computation,''
\newblock {\em {IEEE} Trans. Wireless Commun.}, 2023.

\bibitem{PracticalImplementation3}
A. Sahin,
\newblock ``A demonstration of over-the-air-computation for {FEEL},''
\newblock in {\em IEEE Global Communications Conference Workshops}, 2022, pp. 1--7.

\bibitem{PracticalImplementation4}
L. You, X. Zhao, R. Cao, Y. Shao, and L. Fu,
\newblock ``Broadband digital over-the-air computation for wireless federated edge learning,''
\newblock {\em IEEE Transactions on Mobile Computing}, pp. 1--16, 2023.

\bibitem{sery2021over}
T. Sery, N. Shlezinger, K. Cohen, and Y.~C. Eldar,
\newblock ``Over-the-air federated learning from heterogeneous data,''
\newblock {\em {IEEE} Trans. Signal Process.}, vol. 69, pp. 3796--3811, 2021.

\bibitem{ImperfectDownlink}
W. Guo, R. Li, C. Huang, X. Qin, K. Shen, and W. Zhang,
\newblock ``Joint device selection and power control for wireless federated learning,''
\newblock {\em {IEEE} J. Sel. Areas Commun.}, vol. 40, no. 8, pp. 2395--2410, Aug. 2022.

\bibitem{zhu2018mimo}
G. Zhu, L. Chen, and K. Huang,
\newblock ``{MIMO} over-the-air computation: Beamforming optimization on the grassmann manifold,''
\newblock in {\em Proc. {IEEE} Global Commun. Conf. (GLOBECOM)}, 2018.

\bibitem{jiang2019over}
T. Jiang and Y. Shi,
\newblock ``Over-the-air computation via intelligent reflecting surfaces,''
\newblock in {\em Proc. {IEEE} Global Commun. Conf. (GLOBECOM)}, 2019.

\bibitem{firstyang2020federated}
K. Yang, Y. Shi, Y. Zhou, Z. Yang, L. Fu, and W. Chen,
\newblock ``Federated machine learning for intelligent iot via reconfigurable intelligent surface,''
\newblock {\em IEEE Network}, vol. 34, no. 5, pp. 16--22, 2020.

\bibitem{yang2020federated}
K. Yang, T. Jiang, Y. Shi, and Z. Ding,
\newblock ``Federated learning via over-the-air computation,''
\newblock {\em {IEEE} Trans. Wireless Commun.}, vol. 19, no. 3, pp. 2022--2035, 2020.

\bibitem{Paper2023}
M. Kim, A.~L. Swindlehurst, and D. Park,
\newblock ``Beamforming vector design and device selection in over-the-air federated learning,''
\newblock {\em {IEEE} Trans. Wireless Commun.}, vol. 54, no. 6, pp. 2239--2251, Mar. 2023.

\bibitem{liu2021reconfigurable}
H. Liu, X. Yuan, and Y.-J.~A. Zhang,
\newblock ``Reconfigurable intelligent surface enabled federated learning: A unified communication-learning design approach,''
\newblock {\em {IEEE} Trans. Wireless Commun.}, vol. 20, no. 11, pp. 7595--7609, 2021.

\bibitem{mcmahan2017communication}
B. McMahan, E. Moore, D. Ramage, S. Hampson, and B.~A. y~Arcas,
\newblock ``Communication-efficient learning of deep networks from decentralized data,''
\newblock in {\em Int. Conf. Artificial Intelligence and Statistics}, 2017, pp. 1273--1282.

\bibitem{friedlander2012hybrid}
M.~P. Friedlander and M. Schmidt,
\newblock ``Hybrid deterministic-stochastic methods for data fitting,''
\newblock {\em SIAM Journal on Scientific Computing}, vol. 34, no. 3, pp. A1380--A1405, 2012.

\bibitem{Sidiropoulos&etal:TSP2006}
N.~D. {Sidiropoulos}, T.~N. {Davidson}, and Z.-Q. Luo,
\newblock ``Transmit beamforming for physical-layer multicasting,''
\newblock {\em {IEEE} Trans. Signal Process.}, vol. 54, no. 6, pp. 2239--2251, Jun. 2006.

\bibitem{dong2020multi}
M. Dong and Q. Wang,
\newblock ``Multi-group multicast beamforming: Optimal structure and efficient algorithms,''
\newblock {\em {IEEE} Trans. Signal Process.}, vol. 68, pp. 3738--3753, 2020.

\bibitem{Tranetal:SPL14}
L.-N. Tran, M.~F. Hanif, and M. Juntti,
\newblock ``A conic quadratic programming approach to physical layer multicasting for large-scale antenna arrays,''
\newblock {\em IEEE Signal Processing Letters}, vol. 21, no. 1, pp. 114--117, 2014.

\bibitem{chongTSP23}
C. Zhang, M. Dong, and B. Liang,
\newblock ``Ultra-low-complexity algorithms with structurally optimal multi-group multicast beamforming in large-scale systems,''
\newblock {\em {IEEE} Trans. Signal Process.}, vol. 71, pp. 1626--1641, 2023.

\bibitem{MNIST}
Y. LeCun, C. Cortes, and C. Burges,
\newblock ``The mnist database of handwritten digits,''
\newblock {\em [Online]. Available: http://yann.lecun.com/exdb/mnist/}, 1998.

\bibitem{CIFAR10}
A. Krizhevsky and G. Hinton,
\newblock ``Learning multiple layers of features from tiny images,''
\newblock {\em M.S. thesis, Univ. Toronto, Toronto, ON, Canada, Tech. Rep., [Online]. Available: http://www.cs.toronto.edu/ kriz/learning-features-2009-TR.pdf}, 2009.

\bibitem{ResNet}
K. He, X. Zhang, S. Ren, and J. Sun,
\newblock ``Deep residual learning for image recognition,''
\newblock in {\em Proc. IEEE Conference on Computer Vision and Pattern Recognition}, 2016.

\bibitem{NEURIPS2019_9015}
A. Paszke, S. Gross, F. Massa, A. Lerer, J. Bradbury, G. Chanan, T. Killeen, Z. Lin, N. Gimelshein, L. Antiga, A. Desmaison, A. Kopf, E. Yang, Z. DeVito, M. Raison, A. Tejani, S. Chilamkurthy, B. Steiner, L. Fang, J. Bai, and S. Chintala,
\newblock ``Pytorch: An imperative style, high-performance deep learning library,''
\newblock in {\em Advances in Neural Information Processing Systems 32}, pp. 8024--8035. 2019.

\end{thebibliography}

\end{document}